\documentclass[journal]{IEEEtran}
\usepackage{graphicx,amssymb,amsfonts,amsmath,array}
\usepackage{caption,epsfig,epstopdf}
\usepackage{setspace,exscale,relsize,fancyhdr,float}
\usepackage{cite,textcomp}
\usepackage{algorithm,amsthm}
\usepackage{subfigure}
\usepackage[english]{babel}
\DeclareMathOperator*{\argmin}{argmin}
\newtheorem{theorem}{\textbf{Theorem}}
\newtheorem{lemma}{\textbf{Lemma}}

\begin{document}

\newcommand{\XX}{\mbox{$\boldsymbol{\mathcal{X}}$}}
\newcommand{\YY}{\mbox{$\boldsymbol{\mathcal{Y}}$}}
\newcommand{\ZZ}{\mbox{$\boldsymbol{\mathcal{Z}}$}}
\newcommand{\EE}{\mbox{$\boldsymbol{\mathcal{E}}$}}
\newcommand{\II}{\mbox{$\boldsymbol{\mathcal{I}}$}}
\newcommand{\HH}{\mbox{$\boldsymbol{\mathcal{H}}$}}
\newcommand{\WW}{\mbox{$\boldsymbol{\mathcal{W}}$}}
\newcommand{\FU}{\mbox{$\boldsymbol{\underline{\mathbf{F}}}$}}
\newcommand{\hhat}{\mbox{$\hat{\mathbf{h}}$}}
\newcommand{\Rh}{\mbox{$\mathbf{R}$}}
\newcommand{\N}{\mbox{$\mathcal{N}$}}
\newcommand{\h}{\mbox{$\mathbf{h}$}}

\title{Distributed Channel Estimation and Pilot Contamination Analysis for Massive MIMO-OFDM Systems}

\author{Alam~Zaib, Mudassir~Masood,~Anum~Ali,~Weiyu~Xu and Tareq~Y.~Al{-}Naffouri}
\maketitle

\begin{abstract}
  Massive MIMO communication systems, by virtue of utilizing very large number of antennas, have a potential to yield higher spectral and energy efficiency in comparison with the conventional MIMO systems. In this paper, we consider uplink channel estimation in massive MIMO-OFDM systems with frequency selective channels. With increased number of antennas, the channel estimation problem becomes very challenging as exceptionally large number of channel parameters have to be estimated. We propose an efficient distributed linear minimum mean square error (LMMSE) algorithm that can achieve near optimal channel estimates at very low complexity by exploiting the strong spatial correlations and symmetry of large antenna array elements. The proposed method involves solving a (fixed) reduced dimensional LMMSE problem at each antenna followed by a repetitive sharing of information through collaboration among neighboring antenna elements. To further enhance the channel estimates and/or reduce the number of reserved pilot tones, we propose a data-aided estimation technique that relies on finding a set of most reliable data carriers. We also analyse the effect of pilot contamination on the mean square error (MSE) performance of different channel estimation techniques. Unlike the conventional approaches, we use stochastic geometry to obtain analytical expression for interference variance (or power) across OFDM frequency tones and use it to derive the MSE expressions for different algorithms under both noise and pilot contaminated regimes. Simulation results validate our analysis and the near optimal MSE performance of proposed estimation algorithms.\\


\end{abstract}

\section{Introduction}
In wireless communications, the demand for higher data rates has been dramatically increasing mostly owing to the unprecedented usage of data-hungry devices e.g., smart-phones, super-phones, tablets etc., for wireless multimedia applications \cite{Beyond4G}. Over the years, the MIMO technology (that exploits multiple antennas at the transmitter and/or receiver) has played a pivotal role in sustaining the increased data rates. Installing multiple antennas offers key advantages such as multiplexing gain and diversity gain due to increased spatial reuse \cite{paulraj94,Telatar99}. The MIMO technology has already been incorporated into many wireless products and standards such as WiFi IEEE802.11n\cite{WLAN}, WiMAX IEEE 802.16e\cite{BroadbandOFDM}, LTE (4G) \cite{3GPPLTE}.

Recently, it was established that the use of very large antenna arrays, typically of the order of few hundreds, at the base station (BS) can potentially provide huge gains in system throughput, energy efficiency, security and robustness of wireless communication systems \cite{Marz10}. Such systems, known as massive MIMO or large scale MIMO systems \cite{Rusek13,Larsson14,Debbah13}, overcome many limitations of traditional MIMO systems. Massive MIMO increases system capacity by simultaneously serving tens of users using the same time-frequency resources. Moreover, the large number of low power active antennas allows to focus energy in a small spatial region by forming a sharp beam towards desired users. This additionally implies that there will be little intra-cell interference \cite{Larsson14}. Because of these vital advantages, massive MIMO has attracted a lot of research interest and is envisioned as an enabling technology for next generation (5G) wireless communications \cite{fiveG}.

Hand in hand with the advantages are entirely new research challenges that need to be tackled for massive MIMO. The bottleneck in achieving the full advantages of massive MIMO is the accurate estimation of the channel impulse response (CIR) for each transmit-receive antenna pair. Having a very large number of antennas means that a significant number of channel coefficients need to be estimated $-$ far more than that could be handled by traditional pilot-based MIMO channel estimation techniques (see \cite{Bjornson10} and references therein). In this regard, Bayesian minimum mean square error (MMSE) estimator provides an optimal estimate in the presence of additive white Gaussian noise (AWGN). The method is complex and therefore, a number of approaches have been developed to reduce its complexity such as those proposed in \cite{VBeek95,Edfors96,Ozdemir06,Mehl08,Simko11}. Unlike the least squares (LS) or interpolation based techniques \cite{Ozdemir07}, the MMSE estimation has a clear edge in that it can effectively utilize the channel statistics to improve the estimation accuracy. However, the direct generalization of these techniques to massive MIMO has some drawbacks. In particular, they suffer from huge complexity due to matrix inversion of very large dimensionality, making it impractical. Some methods to reduce the complexity of MMSE estimator in massive MIMO have also been proposed e.g., \cite{Shar13,PengXu14,Gesbert13,HienEVD12,Ghrayeb13,Vetterl11}. It is important to note that most of the existing methods make assumptions that are not always true. For example, many methods deal with flat fading channels only while others assume that the channels are sparse. Therefore, low complexity channel estimation approaches suited to multi-cell and multi-carrier massive MIMO systems need further investigations.

In this paper we propose a distributed algorithm for the estimation of correlated Rayleigh fading channels in massive MIMO-OFDM systems. The novel distributed LMMSE algorithm significantly reduces the computational complexity while attaining near optimal CIR estimates. The distributed approach is inspired by our previous work in \cite{Mudassir15} (where channels are assumed to be sparse and exhibit common support with the neighboring antennas). Furthermore, in order to enhance the estimation performance, we also propose a data-aided estimation technique that relies on finding a set of most reliable data carriers to increase the number of measurements, instead of increasing the reserved pilot tones \cite{Ebrahim12}. Equivalently, by using the data-aided technique, the number of reserved pilot tones can be reduced to attain a performance that is comparable to pilot-based estimation, thus increasing the spectral efficiency.

In a multi-cell setting, allocation of orthogonal pilot sequences for all users cannot be guaranteed due to finite coherence time of the channel and the limited available bandwidth \cite{Marz10}. Therefore, it is inevitable to reuse the pilot sequences across the cells. One of the major consequences of pilots reuse is that when the BS in a cell is performing channel estimation via uplink training, the channel estimates will be severely distorted (contaminated) by the pilots of the neighboring cell users. The impact of pilot contamination on channel estimation is far greater than AWGN. In fact, it was shown in \cite{Larsson11} that the effect of uncorrelated interference and fast Rayleigh fading diminishes as the number of BS antennas increase while the effect of pilot contamination is not eliminated. Hence, it is important to investigate the effect of pilot contamination on MSE performance of different channel estimation techniques. Although the effect of pilot contamination on system performance has been analysed by many researches e.g., \cite{josePC11,Jindal11}, only few studies have analysed its impact on channel estimation performance \cite{PengXu14}. Moreover, in these works, the analysis is carried out for fixed locations of (interference) users. Also it can be seen from analytical expressions derived in these works, that the pathloss, which is determined by user\textquotesingle s locations, plays an important role in MSE performance evaluation.  As such, the above works cannot analytically answer how the randomness of users's locations would effect MSE performance under pilot contamination. In contrast to existing studies, we approach the problem by using concepts from stochastic geometry. By assuming that the interfering users are distributed according to homogeneous poisson point process (PPP), we derive analytical expressions for MSE of LS and LMMSE based channel estimation algorithms in the presence of both AWGN and pilot contamination. The analytical results are validated by simulations. The results clearly show the dependence of important massive MIMO network parameters, such as pathloss and user\textquotesingle s density, on the MSE performance and give clue to mitigate the effect of pilot contamination. It is shown that the increasing the number of pilots does not improve the estimation performance in the presence of pilot contamination. Moreover, the dependence of MSE on antenna spatial correlations suggests that the massive antenna array structure could be optimized to slightly improve the estimation performance under pilot contamination.

The remainder of the paper is organized as follows. Section \ref{sec:sysMod} describes the system and spatial channel correlation model. In Section \ref{sec:LS_LMMSE_est}, we present the MMSE and LS based channel estimation in the presence of AWGN only and discuss their limitations for massive MIMO. The proposed distributed LMMSE algorithm is presented in Section \ref{sec:distEst}. To enhance estimation performance, the data-aided approach is considered in Section \ref{sec:DAest}. Section \ref{sec:PC} describes the effect of pilot contamination on channel estimation, and the expression for interference correlation is presented. Based on this, the MSE expressions for different algorithms are derived under AWGN and pilot contamination. Simulation results are presented in Section \ref{sec:sim} and finally we conclude in Section \ref{sec:concl}.

\subsection{Notations}
We use the lower case letters $x$ and lower case boldface letters $\mathbf{x}$ to represent the scalar and the (column) vector respectively. Matrices are denoted by upper case boldface letters $\mathbf{X}$ whereas the calligraphic notation $\XX$ is reserved for vectors in the frequency domain. The $i$th entry of $\mathbf{x}$ is represented by $x(i)$, the element of $\mathbf{X}$ in $i$th row and $j$th column is denoted by $x_{i,j}$ and the vector $\mathbf{x}_k$ represents the $k$th column of $\mathbf{X}$. We use $\mathbf{x}(\mathcal{P})$ to denote a vector formed by selecting the entries of $\mathbf{x}$ indexed by set $\mathcal{P}$ and $\mathbf{X}(\mathcal{P})$ to denote a matrix formed by selecting the rows of $\mathbf{X}$ indexed by $\mathcal{P}$. We also use $\mathbf{X}_{ij}$ to refer to the $(i,j)$th block entry of a block matrix. Further, $(.)^{\rm T}$, $(.)^*$ and $(.)^{\rm H}$ represent transpose, conjugate and conjugate transpose (Hermitian) operations respectively. We use $\rm{diag}(\mathbf{x})$ to transform a vector $\mathbf{x}$ into a diagonal matrix with the entries of $\mathbf{x}$ spread along the diagonal. $\langle\hat{\XX}(k)\rangle$ denotes the hard decoding i.e., maximum likelihood (ML) decision of $\hat{\XX}(k)$. $\mathbb{E}\{.\}$ represents the statistical expectation. The discrete Fourier transform (DFT) and inverse DFT (IDFT) matrices are represented by $\mathbf{F}$ and $\mathbf{F}^{\rm H}$ respectively, where we the $(l,k)$th entry of $\mathbf{F}$ is defined as $f_{l,k}{=}N^{-1/2}e^{-\jmath2\pi lk/N}$, $l,k{=}0,1,2,\cdots,N-1$ for an $N$-dimensional Fourier transform. Finally, the weighted norm of a vector $\mathbf{x}$ is given by $\|\mathbf{x}\|_{\mathbf{A}}^2\triangleq \mathbf{x}^H\mathbf{A}\mathbf{x}$.

\section{System Model}
\label{sec:sysMod}
\begin{figure}[!t]
 \centering
 \includegraphics[width=0.9\linewidth]{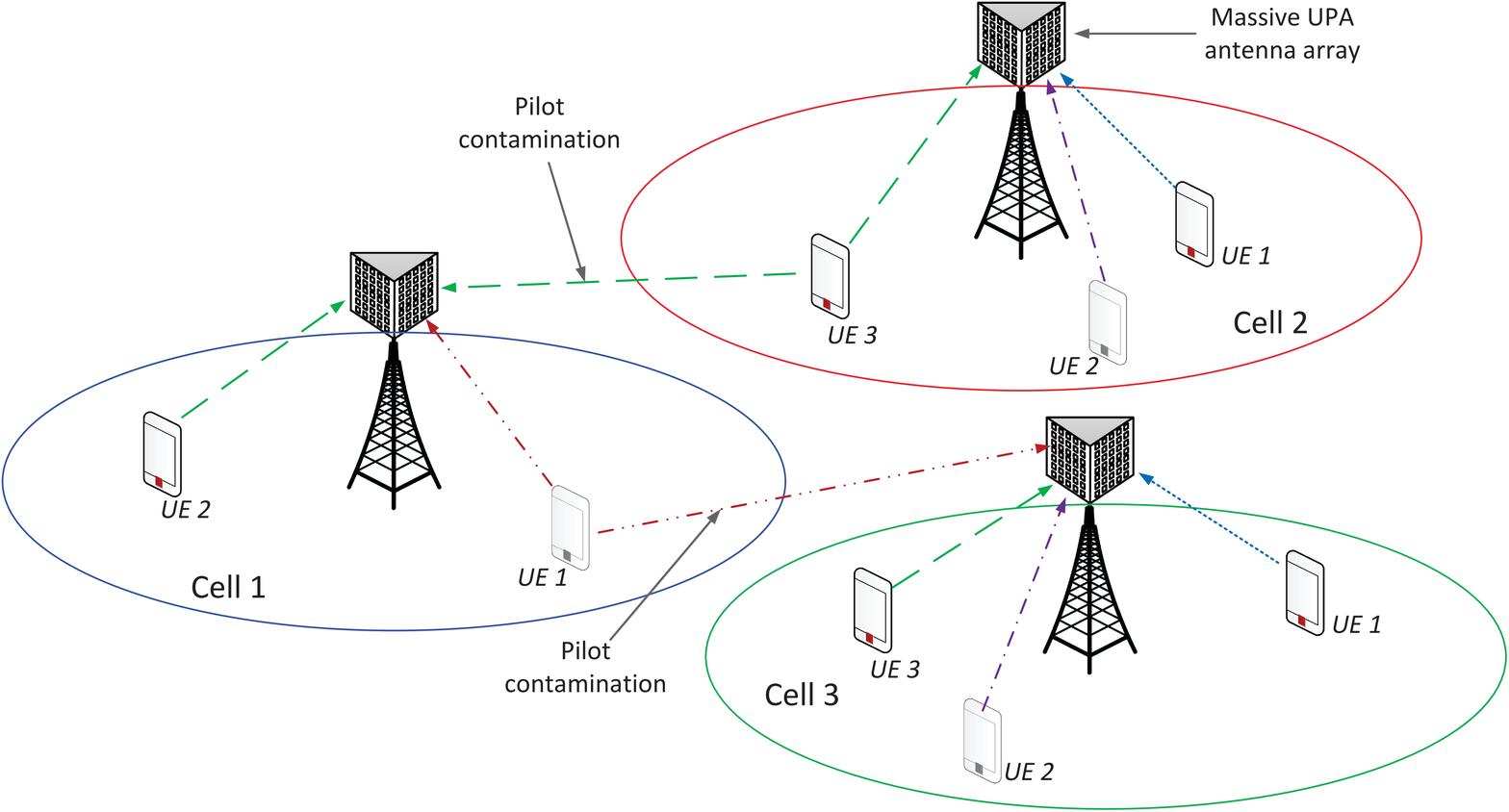}
 \caption{Multi-cell massive MIMO system layout.}
 \label{fig:cell_struct}
\end{figure}

\begin{figure}[!t]
  \centering
  \includegraphics[width=0.7\linewidth]{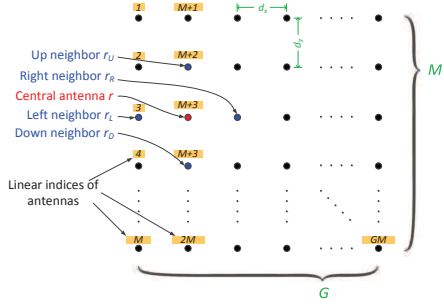}
  \caption{An example of $M\times G$ UPA structure with antenna indexing.}
  \label{fig:UPA_struct}
\end{figure}

We consider a multi-cell massive MIMO-OFDM wireless system as shown in Fig. \ref{fig:cell_struct}, where the BS in each cell is equipped with uniform planar array (UPA) consisting of a large number of antennas. Moreover, we assume that each BS serves a number of single antenna user terminals. The antennas on UPAs are distributed across $M$ rows and $G$ columns with horizontal and vertical spacing of $d_x$ and $d_y$ respectively. We define the $(m,g)$th antenna as the antenna element in $m$th row and $g$th column which corresponds to $r {=}m+M(g-1)$th antenna index  where $1\le m\le M$, $1\le g\le G$ and $1\le r\le R$, where $R{=}MG$ is the total number of antennas in a UPA. Fig. \ref{fig:UPA_struct} shows an example of a $M{\times} G$ UPA structure with antenna indexing. Note that, depending on values of $G$ and $M$, the antennas could have linear or a rectangular configuration. We however, confine our attention to rectangular UPA structure which is a viable configuration in deployment scenarios for massive MIMO \cite{Larsson14}.

Each user communicates with the BS using OFDM and transmits uplink pilots for channel estimation. We assume that all users in a particular cell are assigned orthogonal frequency tones so that there is no intra-cell interference. However, due to necessary reuse of pilots, there are users in the neighboring cells that transmit pilots at the same frequency tones, resulting in an inter-cell interference or pilot contamination. Since only the user in a particular cell of interest will experience interference from the users of neighboring cells that share pilots at the same frequency tones, hence without loss of generality, it suffices to consider one user per cell with all users transmitting pilots at same OFDM frequency tones.

\subsection{Channel Model}
\label{subsec:chanMod}
In the discussion that follows, we assume that there is no inter-cell interference and thus focus on a single-cell single-user scenario (the case of multi-cell will be treated in section \ref{sec:PC} further ahead). Further, we assume a multi-path channel between user and receive antenna $r$ modeled by a Gaussian $L$-tap CIR vector. Specifically, the channel between user and antenna $r$ is defined by $\mathbf{h}_{r} {\triangleq} \left[h_{r}{(0)},h_{r}{(1)},\cdots, h_{r}{(L-1)}\right]^{\rm T}$ where $h_{r}{(l)}\in \mathbb{C}$ represents the $l$th tap complex channel gain. We append all the CIR vectors from a user to the $R$ antennas of the BS to form an $RL$ dimensional composite channel vector $\mathbf{h} {\triangleq} \left[\mathbf{h}_{1}^{\rm T},\mathbf{h}_{2}^{\rm T},\cdots,\mathbf{h}_{R}^{\rm T}\right]^{\rm T}$. Further, we collect the $l$th tap of all transmit-receive pairs to form an $R$ dimensional $l$th tap vector $\mathbf{h}^{(l)}{\triangleq}\left[h_{1}{(l)}, h_{2}{(l)},\cdots,h_{R}{(l)}\right]^{\rm T}$. Then, the $RL\times RL$ dimensional composite channel correlation matrix can be written as,
\begin{equation}\label{eq:Rht}
  \mathbf{R_h} \triangleq \mathbb{E}\{\mathbf{h}\mathbf{h}^{\rm H}\}= \mathbf{R}_{array}\otimes \mathbf{R}_{tap}\:,
\end{equation}
which is the kronecker product ($\otimes$) of two components: (\rm i) The $R{\times} R$ dimensional antenna spatial correlation matrix, $\mathbf{R}_{array} {=}\mathbb{E}\{\mathbf{h}^{(l)}\mathbf{h}^{(l)\rm H}\}, \forall l{=}0,1,\cdots ,L-1$, which represents the correlation among the $l$th taps across the array and (\rm ii) The $L{\times} L$ dimensional channel tap correlation matrix, $\mathbf{R}_{tap}{=}\mathbb{E}\{\mathbf{h}_{r}\mathbf{h}_{r}^{\rm H}\},\forall r{=}1,2,\cdots,R$, which represents the correlation among the CIR taps that depends on channel power delay profile (PDP). As manifested by (\ref{eq:Rht}), $\mathbf{R}_{array}$ is assumed to be identical across the $l$ taps while $\mathbf{R}_{tap}$ is assumed to be identical across the array. For the spatial correlation matrix $\mathbf{R}_{array}$, we adopt a ray-based 3D channel model from \cite{Dawei14} which is more appropriate for rectangular arrays. Accordingly, the spatial correlation between array elements $r{=}(m,g)$ and  $r'{=}(p,q)$ is given by,
\begin{equation}\label{eq:Rs}
  \left[\mathbf{R}_{array}\right]_{r,r'}{=}\frac{D_1}{\sqrt{D_5}}e^{-\frac{D_7+(D_2(sin\phi)\sigma)^2}{2D_5}}e^{\jmath \frac{D_2D_6}{D_5}}\:,
\end{equation}
where the $D_i$\textquotesingle s are defined as,
\begin{align*}
D_1 &= e^{\jmath \frac{2\pi d_x}{\nu}(p-m)cos(\theta)}e^{-\frac{1}{2}(\xi\frac{2\pi d_x}{\nu})^2(p-m)^2sin^2\theta} \nonumber \:,\\
D_2 &= \frac{2\pi d_x}{\nu}(q-g)sin(\theta) \:,\nonumber \\
D_3 &= \xi\frac{2\pi d_x}{\nu}(q-g)cos(\theta) \:,\nonumber \\
D_4 &= \frac{1}{2}\left(\xi\frac{2\pi}{\nu}\right)^2(p-m)(q-g)sin(2\theta) \nonumber \:,\\
D_5 &= (D_3)^2(sin(\phi)\sigma)^2+1 \nonumber \:,\\
D_6 &= D_4(sin(\phi)\sigma)^2+cos(\phi) \nonumber \:,\\
D_7 &= (D_3)^2cos^2\phi-(D_4)^2(sin(\phi)\sigma)^2-2D_4cos\phi \nonumber\:.
\end{align*}
Here, $\nu$ is the carrier-frequency wavelength in meters, $\phi$ and $\theta$ are the mean horizontal angle-of-departure (AoD) and the mean vertical AoD in radians respectively, $\sigma$ and $\xi$ are the standard deviation of horizontal AoD and the standard deviation of vertical AoD respectively.
As shown in \cite{Dawei14}, the spatial correlation matrix can be well approximated as,
\begin{equation}\label{eq:Rs_kron}
\mathbf{R}_{array}\approx \mathbf{R}_{az}\otimes \mathbf{R}_{el}\:,
\end{equation}
where $\mathbf{R}_{az}$ and $\mathbf{R}_{el}$ are the correlation matrices in azimuth (horizontal) and elevation (vertical) directions, having dimensions $(M{\times} M)$ and $(G{\times} G)$ respectively and are defined as,

\begin{align*}\label{eq:RazRel}
  \left[\mathbf{R}_{el}\right]_{m,p} &= e^{\jmath \frac{2\pi d_x}{\nu} (p-m)cos(\theta)}e^{-\frac{1}{2}(\xi\frac{2\pi d_x}{\nu})^2(p-m)^2sin^2\theta} \nonumber\:,\\
  \left[\mathbf{R}_{az}\right]_{g,q} &= \frac{1}{\sqrt{D_5}} e^{-\frac{D_3^2cos^2\phi}{2D_5}} e^{\jmath \frac{D_2cos\phi}{D_5}}e^{-\frac{1}{2}\frac{(D_2\sigma)^2}{D_5}} \nonumber\:.
\end{align*}
\subsection{Signal model}
\noindent We assume that there are $N$ OFDM sub-carriers and let $\XX\:$ represent the $N$-dimensional information symbol whose entries are drawn from a bi-dimensional constellation e.g., Q-QAM. The equivalent time-domain symbol is obtained by taking inverse Fourier transform i.e., $\mathbf{x} {=} \mathbf{F}^{\rm H}\XX$. The time-domain symbol is then transmitted after inserting a cyclic prefix (CP) of length at least $L{-}1$ to avoid inter-symbol-interference (ISI). After removing the CP at the receiver, the frequency-domain OFDM symbol at $r$th antenna can be represented as,
\begin{equation}\label{eq:Yrt}
  \YY_{r} = {\rm diag}(\XX)\HH_{r} + \WW_{r}\:,
\end{equation}
where, $\WW_{r}$ is frequency domain AWGN vector of zero mean and covariance $\mathbf{R}_w{=}\sigma_w^2\mathbf{I}_N$ and $\HH_{r}$ is the channel frequency response between the user and receive antenna $r$ i.e.,
\begin{equation}\label{eq:CFR}
  \HH_{r} = \sqrt{N}\mathbf{F}\left[ \begin{array}{c}
                        \mathbf{h}_{r} \\
                        \mathbf{0}_{N-L\times 1} \\
                        \end{array} \right] = \sqrt{N}\FU\mathbf{h}_{r}\:.
\end{equation}
Here $\FU$ is truncated Fourier matrix formed by selecting the first $L$ columns of $\mathbf{F}$. Using (\ref{eq:CFR}), we can re-write (\ref{eq:Yrt}) as,
\begin{equation}\label{eq:Yrt2}
  \YY_{r} = \sqrt{N}{\rm diag}(\XX)\FU \mathbf{h}_{r} + \WW_{r} = \mathbf{A}\mathbf{h}_{r} + \WW_{r}\:,
\end{equation}
where $\mathbf{A}{\triangleq} \sqrt{N}{\rm diag}(\XX)\FU$ and the noise vector $\WW_{r}$ is assumed to be uncorrelated with the channel vector $\mathbf{h}_{r}$. We assume that $K$ sub-carriers are reserved for pilots and the remaining $N-K$ for the data transmission. Further, it is best to allocate the pilots uniformly as shown in \cite{Cioffi98}. Hence, for a set of pilot indices denoted by vector $\mathcal{P}$, the system equation (\ref{eq:Yrt2}) reduces to,
\begin{equation}\label{eq:Yrt_pilots}
  \YY_{r}(\mathcal{P}) = \mathbf{A}(\mathcal{P})\mathbf{h}_{r} + \WW_{r}(\mathcal{P})\:,
\end{equation}
where $\YY_{r}(\mathcal{P})$ and $\WW_{r}(\mathcal{P})$ are formed by selecting the entries of $\YY_{r}$ and $\WW_{r}$ indexed by $\mathcal{P}$ while $\mathbf{A}(\mathcal{P})$ is a $K\times L$ matrix formed by selecting the rows of $\mathbf{A}$ indexed by $\mathcal{P}$.

We can now collect the pilot measurements (\ref{eq:Yrt_pilots}) received by all antennas into a single system of equations as follows,
\begin{equation}\label{eq:Y_pilots}
 \YY(\mathcal{P})=[\mathbf{I}_{R}\otimes \mathbf{A}(\mathcal{P})]\mathbf{h}+\WW(\mathcal{P})\:,
\end{equation}
where, $\YY(\mathcal{P}){=}\left[\YY_1^{\rm T}(\mathcal{P}),\cdots,\YY_{R}^{\rm T}(\mathcal{P})\right]^{\rm T}$, $\WW(\mathcal{P})=\left[\WW_1^{\rm T}(\mathcal{P}), \cdots,\WW_{R}^{\rm T}(\mathcal{P})\right]^{\rm T}$, $\mathbf{I}_{R}$ represents an $R\times R$ identity matrix and $\mathbf{h}$, as defined earlier, represents the composite channel vector from user to the BS. For convenience, we assume the noise variance to be identical across the array so that $\WW(\mathcal{P})\sim \mathcal{CN} (\mathbf{0},\mathbf{R}_w{=}\sigma_w^2\mathbf{I}_{RK})$. Note that the number of unknown channel coefficients in (\ref{eq:Y_pilots}) are $RL$ whereas the total number of equations are $RK$. Therefore, a necessary condition to solve (\ref{eq:Y_pilots}) for $\mathbf{h}$ (and also (\ref{eq:Yrt_pilots}) for $\mathbf{h}_r$) using least squares, is that the number of pilots be at least equal to $L$ i.e., $K\ge L$. However, $K$ could be reduced if we utilize the correlation information. With the models defined above, we are ready to estimate the CIRs between the user and each BS antenna. We pursue different approaches that can be adopted for channel estimation in massive MIMO setup depending on whether the information processing takes place independently at each antenna element or jointly at a centralized processor.
We start with naive LMMSE and LS based techniques and discuss their limitations, and then propose a new distributed  approach in section \ref{sec:distEst} which is further extended in section \ref{sec:DAest} with the help of data-aided approach.
\section{LMMSE and LS based Channel Estimation}
\label{sec:LS_LMMSE_est}
 In this section, we present three different techniques for channel estimation in massive MIMO-OFDM based on the well-known LMMSE and LS estimators and discuss their limitations. For now, we assume that estimates are corrupted only by the white noise. Hence, without loss of generality, we consider a single-cell single-user scenario for the approaches presented below.
\subsection{The Localized LMMSE (L-LMMSE) estimation}
\label{subsec:LLMMSE_est}
In this approach, all CIRs are estimated independently based on the observations received at each antenna element by using the classical LMMSE estimation. Using the linear system model in (\ref{eq:Yrt_pilots}), the LMMSE estimate of $\mathbf{h}_r$ is obtained by minimizing the (local) MSE, $\mathbb{E}\{\|\mathbf{h}_r{-}\hat{\mathbf{h}}_r\|^2\}$, over $\hat{\mathbf{h}}_r$ as follows \cite{sayed03}
\begin{equation}\label{eq:hhatSep}
  \hat{\mathbf{h}}_r = \left(\mathbf{R}_{tap}^{-1}+ \mathbf{A}^{\rm H}\mathbf{R}_w^{-1}\mathbf{A}\right)^{-1}\mathbf{A}^{\rm H}\mathbf{R}_w^{-1}\YY_r\:,
\end{equation}
where we drop the index vector $\mathcal{P}$ for convenience. Similarly, it follows that the (minimum) MSE  is,
\begin{equation}\label{eq:MSESep}
  {\rm mse}_r = {\rm trace}\left(\mathbf{R}_{tap}^{-1}+ \mathbf{A}^{\rm H}\mathbf{R}_w^{-1}\mathbf{A}\right)^{-1}\:.
\end{equation}
The overall global MSE is obtained by taking summation over all array elements i.e., ${\rm MSE^{(L)}} {=} \sum_{r=1}^{R} {\rm mse}_r$, which after simplifying (\ref{eq:MSESep}), can be expressed as,
\begin{equation}\label{eq:MSESep_simple}
  {\rm MSE^{(L)}} = R\sum_{i=1}^L\left(\frac{\delta_i}{1+\rho K\delta_i}\right)\:,
\end{equation}
where $\{\delta_i\}_{i=1}^L$ are eigenvalues of $\mathbf{R}_{tap}$, $\rho\triangleq E_x/\sigma^2_w$ is the SNR with $E_x$ representing the average signal energy per symbol and the superscript $({\rm L})$ indicates L-LMMSE. Observe from (\ref{eq:MSESep_simple}) that channel delay spread $L$, has an adverse effect on MSE performance, which can be reduced by increasing the number of pilot tones. The computational complexity of L-LMMSE is of the order $O\big(RL^3\big)$ (see Table \ref{tab:comp}), which increases linearly with the number of BS antennas. However, the CIR estimates are not optimal in the sense of minimizing the overall or global MSE.  The estimates would have been optimal, had the antennas been placed sufficiently apart so that the channel vectors were effectively uncorrelated. But for massive MIMO with extremely large number of antennas, it is expected that antennas are located in close proximity, so the channel vectors are highly likely to be correlated with each other.

\subsection{The Optimal LMMSE (O-LMMSE) Solution}
\label{subsec:OLMMSE_est}
In this strategy all the channel vectors are estimated simultaneously by minimizing the global MSE, $\mathbb{E}\{\|\mathbf{h}-\hat{\mathbf{h}}\|^2\}$ over the composite channel vector $\hat{\mathbf{h}}$. This could be realized by sending all observations to a central processor and then invoking the LMMSE estimation based on the composite system model in (\ref{eq:Y_pilots}). The solution to this problem is given by,
\begin{equation}\label{eq:hhatComb}
  \hat{\mathbf{h}}^{\rm} = \left(\mathbf{R}_{\mathbf{h}}^{-1}+\acute{\mathbf{A}}^{\rm H}\mathbf{R}_w^{-1}\acute{\mathbf{A}}\right)^{-1} \acute{\mathbf{A}}^{\rm H}\mathbf{R}_w^{-1}\YY\:,
\end{equation}
where, $\acute{\mathbf{A}}{=}\mathbf{I}_{R}\otimes \mathbf{A}$, $\mathbf{R}_{\mathbf{h}}$ is as given in (\ref{eq:Rht}) and for notational convenience we dropped the index $\mathcal{P}$. The corresponding MSE is,
\begin{equation}\label{eq:MSEOpt}
  {\rm MSE^{(O)}}={\rm trace}\left(\mathbf{R}_{\mathbf{h}}^{-1}+ \acute{\mathbf{A}}^{\rm H}\mathbf{R}_w^{-1}\acute{\mathbf{A}}\right)^{-1}\:,
\end{equation}
which can be simplified to yield,
\begin{equation}\label{eq:MSEOpt_simple}
  {\rm MSE^{(O)}} = \sum_{j=1}^{R}\sum_{i=1}^{L}\frac{\eta_j\delta_i}{1+\rho K\eta_j\delta_i}\:,
\end{equation}
where, $\eta_j$ and $\delta_i$ are eigenvalues of $\mathbf{R}_{array}$ and $\mathbf{R}_{tap}$ respectively. By comparing (\ref{eq:MSEOpt_simple}) with (\ref{eq:MSESep_simple}), we conclude that in presence of spatial correlation, the optimal solution yields better MSE performance than the localized strategy, however, it has the following two major drawbacks:
\begin{enumerate}
\item Realization of optimal strategy requires global sharing of information to/from the central processor that results in communication overhead (as it requires complex signalling which can be very expensive).
\item As evident from (\ref{eq:hhatComb}), the computation of optimal LMMSE requires inverting a non-trivial matrix of very high dimension ($RK\times RK$) that leads to computational complexity of order $O\big(R^3L^3\big)$, which is cubic in number of BS antennas.
\end{enumerate}
In massive MIMO scenario where $R$ is of the order of few hundreds, both of the above mentioned operations are very expensive and possibly impractical.
\subsection{Estimation using Least Square (LS)}
\label{subsec:LS_est}
If the channel statistics are unknown, one can employ simple LS based estimation. In the absence of  correlation, we can let the inverse of channel correlation matrix go to zero, i.e.,  $\mathbf{R}_{tap}^{-1}\rightarrow\mathbf{0}$, thereby ignoring the channel statistics. Therefore, the localized LS solution from $(\ref{eq:hhatSep})$ is,
\begin{equation}\label{eq:hhatSepLS}
  \hat{\mathbf{h}}_r^{\rm ls} = \left(\mathbf{A}^H\mathbf{A}\right)^{-1}\mathbf{A}^H\YY_r\:,
\end{equation}
and the resulting MSE is given by,
\begin{equation}\label{eq:mseSepLS}
  {\rm mse}_r^{\rm ls} = {\rm trace}\left(\mathbf{A}^{\rm H}\mathbf{R}_w^{-1}\mathbf{A}\right)^{-1}\:.
\end{equation}
In this case, the overall MSE simplifies to,
\begin{equation}\label{eq:mseSepLS_simple}
  {\rm MSE}^{\rm(LS)}=\sum_{r=1}^{R} {\rm mse}_r^{\rm ls}=\frac{RL}{\rho K}\:.
\end{equation}
Comparing (\ref{eq:mseSepLS_simple}) with (\ref{eq:MSESep_simple}), we conclude that LS has poor performance in comparison with the LMMSE as it does not utilize the channel statistics. It is for this reason that the centralized LS (C-LS) solution would achieve the same MSE performance as the localized one as shown below.
\begin{align*}\label{eq:mseCombLS}
  {\rm MSE}^{\rm (C-LS)} &{=} {\rm trace}\left(\left(\mathbf{I}_{R}\otimes\mathbf{A}\right)^{\rm H}\left(\mathbf{I}_{R}\otimes\mathbf{R}_w\right)^{-1} \left(\mathbf{I}_{R}\otimes\mathbf{A}\right)\right)^{-1}\nonumber\\
  &= {\rm trace}\left(\mathbf{I}_{R}\otimes\mathbf{A}^{\rm H}\mathbf{R}_w^{-1} \mathbf{A}\right)^{-1}\nonumber\:,\\
  &= \sum_{r=1}^{R} {\rm trace}\left(\mathbf{A}^{\rm H}\mathbf{R}_w^{-1}\mathbf{A}\right)^{-1}\nonumber\:,\\
  &= {\rm MSE}^{\rm(LS)}\:,
\end{align*}
where we have used the Kronecker product identities, $(\mathbf{A}\otimes\mathbf{B})(\mathbf{C}\otimes\mathbf{D}){=} \mathbf{AC}\otimes\mathbf{BD}$ and $(\mathbf{A}\otimes \mathbf{B})^{-1}{=}\mathbf{A}^{-1}\otimes \mathbf{B}^{-1}$.

In short, the L-LMMSE estimation has the advantage of low complexity (and better performance than LS) but it is unable to exploit the strong spatial correlation among antenna elements which is inevitable in massive MIMO systems. On the other hand, O-LMMSE exploits the spatial correlations but at a significantly higher computational cost. This motivates us to propose a method that can overcome the shortcomings of aforementioned techniques without affecting the estimation quality. Specifically, we propose a distributed estimation of  CIRs based on antenna coordination that attains near optimal performance with tractable complexity. The  proposed distributed LMMSE estimation is described below and is further extended in section \ref{sec:DAest} via a data-aided technique.

\section{The Proposed distributed LMMSE (D-LMMSE) estimation}
\label{sec:distEst}
It is well known from equivalence results in linear estimation theory \cite{kailath} that the O-LMMSE solution (\ref{eq:hhatComb}) could be alternatively obtained by solving an $RL$ dimensional optimization problem,
\begin{equation}\label{eq:OptCost2}
   \argmin_{\mathbf{h}}\left\|\YY-\mathbf{A}'\mathbf{h}\right\|_{\mathbf{R}_w^{-1}}^2    + \left\|\mathbf{h}\right\|_{\mathbf{R}_{\mathbf{h}}^{-1}}^2\:,
\end{equation}
% = \argmin_{\mathbf{h}_1,\cdots,\mathbf{h}_{R}} \sum_r\left\|\YY_r-\mathbf{A}\mathbf{h}_r\right\|_{\mathbf{R}_w^{-1}}^2    + \left\|\mathbf{h}\right\|_{\mathbf{R}_{\mathbf{h}}^{-1}}^2
where all the variables are as defined earlier. Instead of solving (\ref{eq:OptCost2}) globally (as done earlier), we aim to solve it in a distributed manner over $R$ antennas in which the $r$th antenna has access to $\YY_r$ only. Moreover, the antenna $r$ is interested only in determining its own CIR (i.e., $\mathbf{h}_r$) without worrying about other $\mathbf{h}_j$\textquotesingle s. Here, we would like to mention that this problem is fundamentally different from those considered in the context of adaptive networks \cite{sayed14}. Also, most of the existing distributed estimation techniques in adaptive networks deal with single task problems in which all nodes in the network estimate a single common parameter of interest. Furthermore, they rely on full cooperation between the nodes, i.e., exchanging both the estimates and the observations with the neighbors. Although, the distributed recursive least squares (RLS) algorithm of \cite{DRLS08} might be adopted to solve (\ref{eq:OptCost2}), it would be gravely complex in number of dimensions (due to large $R$ in massive MIMO and large channel delay spread) and hence might suffer from convergence issues. Our proposed solution, the distributed LMMSE (D-LMMSE) algorithm, as will become clear, is much simpler in that it exploits the structure of spatial correlation matrix $\mathbf{R}_{array}$ and relies only on exchanging the (partial) weighted estimates of CIRs with immediate neighbors, thus reducing the communication and computational cost significantly. The proposed D-LMMSE algorithm is composed of three main steps namely the estimation, sharing and update, as explained below.

\subsection{Estimation}
\label{subsec:EstStep}
In the estimation step, each antenna acting as a center antenna $r_C$, estimates not only its own CIR but also the CIRs of its neighborhood. The neighborhood of $r_C$ consists of 4-direct neighbors represented by the set $\mathcal{N}{=}\{r_L, r_R, r_U,r_D\}$\footnote{Note that for elements lying at the edges of a UPA, the number of neighbors are different, so that $2\le |\mathcal{N}|\le 4$. The set of neighbors including the central antenna is represented by $\mathcal{N}^+$.} on the left, right, top and bottom positions respectively as shown in Fig. \ref{fig:infoshare1}. Also, let the corresponding channel vectors be represented by $\mathbf{h}_C$, $\mathbf{h}_L$, $\mathbf{h}_R$, $\mathbf{h}_U$ and $\mathbf{h}_D$ respectively and let $\mathbf{h}^{c}$ represent $|\mathcal{N}^+|L\times 1$ dimensional composite channel vector of the central antenna and its $|\mathcal{N}|$ direct neighbors (i.e., $\mathbf{h}^{\rm c}{=}\left[\mathbf{h}_C^{\rm T},\mathbf{h}_L^{\rm T},\mathbf{h}_R^{\rm T},\mathbf{h}_U^{\rm T},\mathbf{h}_D^{\rm T}\right]^{\rm T}$). During the estimation process, each antenna acting as a central antenna computes the estimate of $\mathbf{h}^{c}$ by solving a reduced dimensional weighted least squares (WLS) optimization problem,
\begin{equation}\label{eq:estObj}
   \hat{\mathbf{h}}^{c} = \argmin_{\mathbf{h}^{c}}\left\|\YY_C(\mathcal{P})-\mathbf{A}(\mathcal{P})\mathbf{h}_C\right\|_{\mathbf{R}_w^{-1}}^2    + \left\|\mathbf{h}^{c}\right\|_{\mathbf{R}_{\mathbf{h}^{c}}^{-1}}^2\:,
\end{equation}
where $\YY_C(\mathcal{P})$ represents pilot observations at the central antenna, $\mathbf{R}_{\mathbf{h}^{c}}$ is channel correlation matrix defined as $\mathbf{R}_{\mathbf{h}^{c}} \triangleq \mathbb{E}\{\mathbf{h}^{c}(\mathbf{h}^{c})^{\rm H}\}$ and $\mathbf{R}_w{=}\sigma_w^2\mathbf{I}_K$ is the noise covariance matrix at the central antenna. From (\ref{eq:estObj}) it is clear that information is processed locally at each antenna as each antenna uses only its own observations and interacts with its neighborhood only through $\mathbf{R}_{\mathbf{h}^{c}}$ (it is assumed that the central antenna has available correlation information of its neighborhood to construct $\mathbf{R}_{\mathbf{h}^{c}}$). The solution to the above WLS minimization problem can be obtained by first re-writing (\ref{eq:estObj}) explicitly in terms of $\mathbf{h}^{c}$ as,
\begin{equation}\label{eq:estObj2}
   \hat{\mathbf{h}}^{c} = \argmin_{\mathbf{h}^{c}}\left\|\bar{\YY}-\bar{\mathbf{A}}\mathbf{h}^{c} \right\|_{\mathbf{R}_w^{-1}}^2 + \left\|\mathbf{h}^{c}\right\|_{\mathbf{R}_{\mathbf{h}^{c}}^{-1}}^2\:,
\end{equation}
where, $\bar{\YY}{=}\YY_C(\mathcal{P})$ and $\bar{\mathbf{A}}{=}\begin{bmatrix}\mathbf{A}(\mathcal{P}) & \mathbf{0}_{K\times L|\mathcal{N}|} \end{bmatrix}$. Then, by invoking the equivalence between LMMSE and WLS estimation problems we obtain,
\begin{equation}\label{eq:estSol}
  \hat{\mathbf{h}}^{c} = \left(\mathbf{R}_{\mathbf{h}^{c}}^{-1}+\bar{\mathbf{A}}^{\rm H}\mathbf{R}_w^{-1} \bar{\mathbf{A}}\right)^{-1}\bar{\mathbf{A}}^{\rm H}\mathbf{R}_w^{-1}\bar{\YY}\:.
\end{equation}
We define $\mathbf{P}^{c}\triangleq\left(\mathbf{C}_e^{c}\right)^{-1}$ as the inverse of error covariance matrix at the central element, which is given by the expression,
\begin{equation}\label{eq:estCe}
  \mathbf{P}^{c} = \mathbf{R}_{\mathbf{h}^{c}}^{-1}+\bar{\mathbf{A}}^{\rm H}\mathbf{R}_w^{-1}\bar{\mathbf{A}}\:.
\end{equation}
Then, by using (\ref{eq:estCe}) into (\ref{eq:estSol}), the weighted estimate of composite channel at each antenna is simply,
\begin{equation}\label{eq:wtEstSol}
  \hat{\mathbf{h}}_w^{c} = \mathbf{P}^c\hat{\mathbf{h}}^{c} = \bar{\mathbf{A}}^{\rm H}\mathbf{R}_w^{-1}\bar{\YY}\:.
\end{equation}
This weighting of the estimates asserts that we put more confidence into the estimates which are more reliable and vice versa.  The estimation step is non-recursive and is computed once for all antennas in the array.  Thus, having found the $\mathbf{P}$ matrices in (\ref{eq:estCe}) and the weighted estimates in (\ref{eq:wtEstSol}), each antenna is ready to initiate sharing.

\subsection{Sharing}
\label{subsec:shareStep}
\begin{figure*}[!t]
 \centerline{\subfigure[Information diffusion process]{
   \includegraphics[width=0.7\columnwidth]{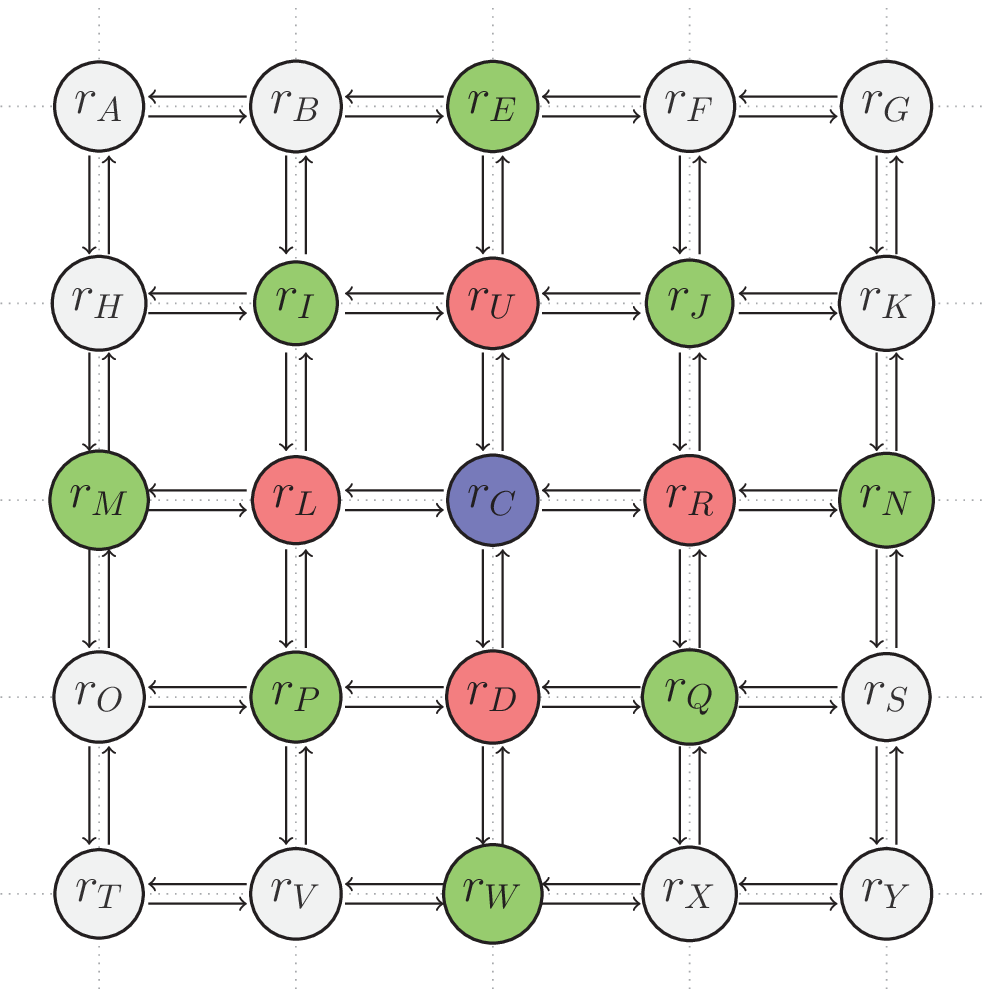}
   \label{fig:infoshare1}}
\hfill
 \subfigure[Information sharing process]{
   \includegraphics[width=0.7\columnwidth]{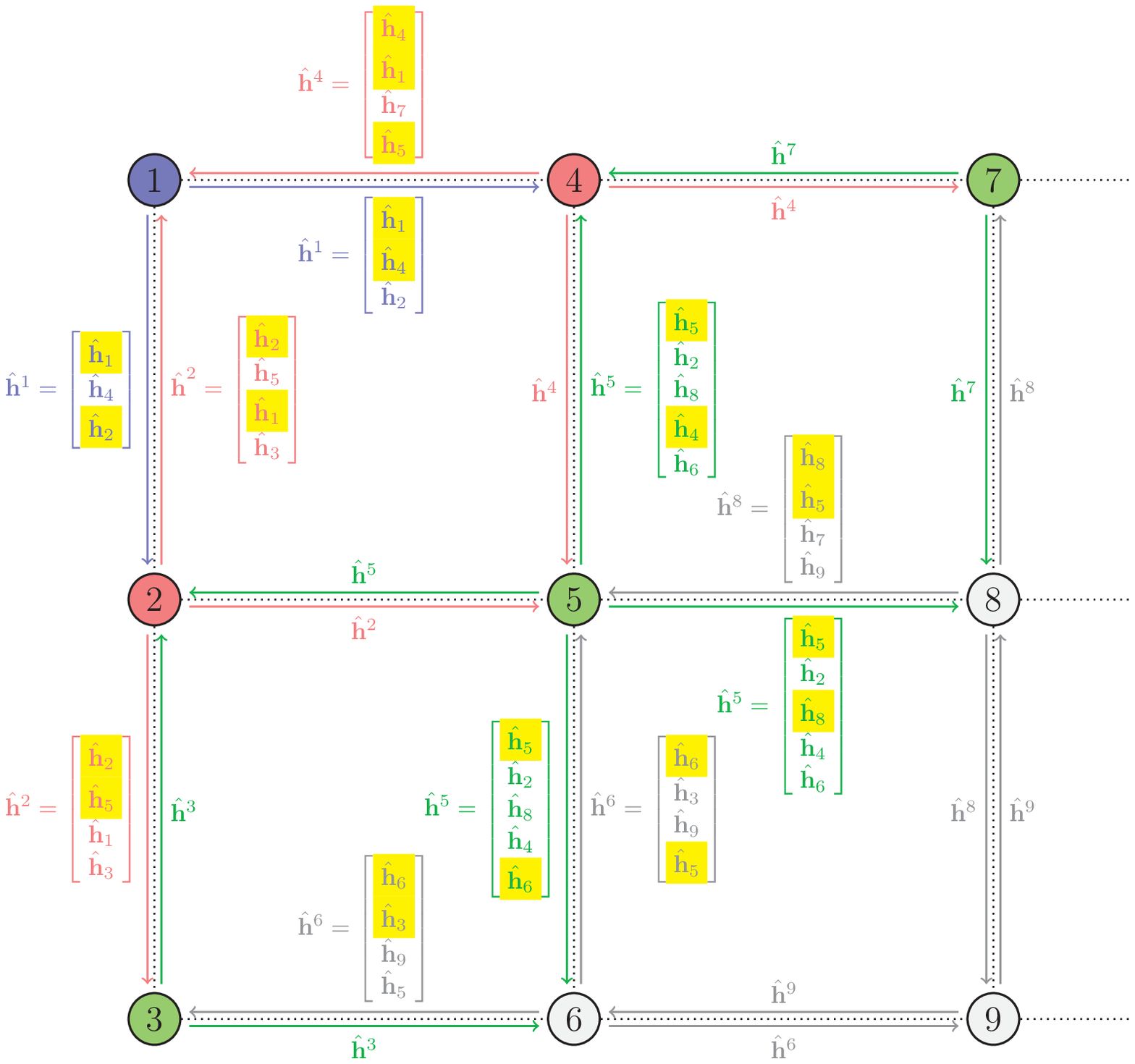}
   \label{fig:infoshare2}}}
\caption{\subref{fig:infoshare1} During the first iteration $r_C$ (blue antenna) receives information from its 4-direct neighbors (pink antennas). In the second iteration, the information from next nearest neighbors (green antennas) also comes in and so on. \subref{fig:infoshare2} An example of a $3\times 4$ antenna array where the neighboring antennas (indices $4$ and $2$) share the selected estimates (highlighted) with the central antenna (index $1$).}
%  \label{fig:infoshare}
\end{figure*}

The sharing step is the key to the proposed distributed algorithm where the information is shared through collaboration between antennas. Let us define the sub-vector $\hhat_{wj}$ of composite vector $\hhat_w^k$ as a (weighted) CIR estimate of antenna $j$ (i.e. the vector $\hhat_j$) computed by the antenna $k$. In sharing step, each antenna acting as a central element, shares only the partial information with its neighbors such that the antenna $k$, with composite vector $\hhat_w^k$, would share only the selected components; its own (weighted) estimate $\hhat_{wk}$  and the (weighted) estimate $\hhat_{wj}$, $j\in\mathcal{N}$, with its $j$th neighbor. Henceforth, the shared vectors will be termed as partial vectors and represented by an underlined notation. An example of how this sharing takes place is also depicted in Fig. \ref{fig:infoshare2} for a $3\times 4$ array with central element $r_C{=}1$ having only two neighbors; $\mathcal{N}{=}\{r_R{=}4,r_D{=}2\}$. As shown, each of the neighboring element shares only two sub-vectors (i.e., partial information) of its composite vector with the central antenna. The collaboration between the rest of the array elements takes place in a similar fashion.

 As a result of information sharing, each antenna acting as a central node $r_C$ receives $|\N|$ partial vectors, $\underline{\hhat}_w^j,j\in\N$, from its neighbors, each of dimension $|\N^+|L\times 1$ and having only two non-zero components; $\hhat_{wj}$ and $\hhat_{wc}$. For the example in Fig. \ref{fig:infoshare2}, the composite vector of the central node and the partial vectors received from its neighbors are given as follows,
\begin{equation}\label{eq:h_Reordered}
  \hat{\mathbf{h}}_w^1 = \begin{bmatrix} \hat{\mathbf{h}}_{w1} \\
                          \hat{\mathbf{h}}_{w4} \\
                          \hat{\mathbf{h}}_{w2} \\
                       \end{bmatrix} \text{,}\:\:\:
  \underline{\hat{\mathbf{h}}}_w^4 = \begin{bmatrix} \hat{\mathbf{h}}_{w1} \\
                          \hat{\mathbf{h}}_{w4} \\
                          \mathbf{0} \\
                       \end{bmatrix}\:\:\text{and}\:\:\:
  \underline{\hat{\mathbf{h}}}_w^2 = \begin{bmatrix} \hat{\mathbf{h}}_{w1} \\
                          \mathbf{0} \\
                          \hat{\mathbf{h}}_{w2} \\
                       \end{bmatrix}\:.
\end{equation}
Note that the estimates which are not shared have been assigned as null vectors.

\subsection{Update}\label{subsec:update}
Having received the (partial) LMMSE estimates from the neighboring elements, each antenna acting as the central element updates its estimate and error covariance matrix. The update rule is based on the optimal combining of estimators, a standard result in LMMSE estimation theory. The result is summarized in the following lemma,

\begin{lemma}\label{lemma:update}
Let $\mathbf{y}_1$ and $\mathbf{y}_2$ be two separate observations of a zero mean random vector $\mathbf{h}$, such that $\mathbf{y}_1{=}\mathbf{A}_1\mathbf{h}+\mathbf{w}_1$ and $\mathbf{y}_2{=}\mathbf{A}_2\mathbf{h}+\mathbf{w}_2$, where we assume that $\mathbf{h}$ is uncorrelated with both $\mathbf{w}_1$ and $\mathbf{w}_2$. Let $\hat{\mathbf{h}}_1$ and $\hat{\mathbf{h}}_2$ denote the LMMSE estimates of $\mathbf{h}$ and $\mathbf{C}_1$ and $\mathbf{C}_2$ be the corresponding error covariance matrices in two experiments. Then, the optimal LMMSE estimator and the error covariance matrix of $\mathbf{h}$ given both the observations are,
\begin{equation}\label{eq:lemma1}
   \mathbf{C}^{-1}\hat{\mathbf{h}} = \mathbf{C}_1^{-1}\hat{\mathbf{h}}_1 + \mathbf{C}_2^{-1}\hat{\mathbf{h}}_2\:,
\end{equation}
and
\begin{equation}\label{eq:lemma2}
   \mathbf{C}^{-1} = \mathbf{C}_1^{-1} + \mathbf{C}_2^{-1}+\mathbf{R}_h^{-1}-\mathbf{R}_1^{-1}-\mathbf{R}_2^{-1}\:,
\end{equation}
where, $\mathbf{R}_h{=}\mathbb{E}\{\mathbf{h}\mathbf{h}^{\rm H}\}$ and $\mathbf{R}_1$ and $\mathbf{R}_2$ are covariance matrices of $\mathbf{h}$ in two experiments.
\end{lemma}

\begin{IEEEproof}
  See \cite{kailath}.
 \end{IEEEproof}

\noindent Aforementioned lemma can be easily extended to more than two observations. The lemma suggests an optimal way of combining the individual estimates obtained by independent observations. We use this lemma at each antenna to improve the initial channel estimate by combining it with the estimates computed and shared by $|\mathcal{N}|$ neighbors. Consequently, by treating each antenna as a central element $r_C$, the update rule is given by following equations,
\begin{equation}\label{eq:estUpdate}
  \hat{\mathbf{h}}_w^{{c}(i)} = \hat{\mathbf{h}}_w^{{c}(i-1)}+\sum_{j\in\mathcal{N}} \hat{\underline{\mathbf{h}}}_w^{{j}(i-1)}\:,
\end{equation}
and
\begin{equation}\label{eq:CeUpdate}
  \mathbf{P}^{c(i)} = \mathbf{P}^{c(i-1)}+\sum_{j\in\mathcal{N}} \left(\underline{\mathbf{P}}^{j(i-1)}-\underline{\mathbf{R}}_{\mathbf{h}^j}^{-1}\right)\:,
\end{equation}
where $\underline{\mathbf{P}}^j$ and $\underline{\mathbf{R}}_{\mathbf{h}^j}$ represent the partial (inverse) error covariance and correlation matrices associated with the partial estimates $\hat{\underline{\mathbf{h}}}_w^j$ and $i$ represents the iteration index. Note that in the update equations, we  employed the weighted estimates and inverse error covariance matrices to minimize the computational requirements. The recursions in the update equations are initialized by (\ref{eq:wtEstSol}) and (\ref{eq:estCe}) respectively, which are available after the estimation step. In the subsequent iterations, each antenna would also require the partial matrices, $\underline{\mathbf{P}}^j$\textquotesingle s and $\underline{\mathbf{R}}_{\mathbf{h}^j}$\textquotesingle s,  for each of its $|\mathcal{N}|$ neighbors. Fortunately, they can be obtained from $\mathbf{P}^{c}$ and $\mathbf{R}_{\mathbf{h}^c}$ respectively (which are available at the central antenna) by exploiting the symmetrical structure of $\mathbf{R}_{array}$. Thus, there is no need to share them across the neighboring elements, that in turn saves a significant amount of communication burden.
Specifically, the matrices $\mathbf{R}_{\mathbf{h}^c}$ and $\mathbf{P}^{c}$ exhibit the following two properties:\footnote{These properties are generally satisfied as the spatial correlation matrix is usually symmetric, if not, then the antennas can share these matrices as well.}\\

\noindent\emph{Property 1:} The matrix $\mathbf{R}_{\mathbf{h}^c}$ is identical for all elements in the neighborhood of $r_C$ i.e., $\mathbf{R}_{\mathbf{h}^c}{=}\mathbf{R}_{\mathbf{h}^j},\forall j\in \mathcal{N}$\\
\noindent\emph{Property 2:} The matrix $\mathbf{P}^c$ is identical for all elements in the neighborhood of $r_C$ i.e., $\mathbf{P}^c{=}\mathbf{P}^j,\forall j\in \mathcal{N}$\\

\noindent Property 1 is attributed to the symmetric nature of the spatial correlation matrix $\mathbf{R}_{array}$, which implies that the spatial correlation between any two antennas, placed equidistant apart, is the same. Therefore, it is not difficult to see that property 1 holds exactly under the Kronecker model and our earlier assumption of identical tap correlation across the antenna array in section \ref{sec:sysMod}. Property 2 is the consequence of property 1 when incorporated into (\ref{eq:estCe}).

Hence, to obtain the patrial correlation matrices, $\underline{\mathbf{R}}_{\mathbf{h}^j},j\in \mathcal{N}$, we use property 1 to first set $\underline{\mathbf{R}}_{\mathbf{h}^j}{=}\mathbf{R}_{\mathbf{h}^c}$ and then modify the off-diagonal block entries corresponding to the null vectors of partial estimates as $\underline{\mathbf{R}}_{ij}{=}\mathbf{0}$ if any $\hat{\mathbf{h}}_{wi},\hat{\mathbf{h}}_{wj}{=}\mathbf{0}$ and the diagonal block entries as $\underline{\mathbf{R}}_{ii}{=}\mathbf{I}_L$ if $\hat{\mathbf{h}}_{wi}{=}\mathbf{0}$, where the subscript $ij$ denotes the $(i,j)$th block. The matrices $\underline{\mathbf{P}}^j$\textquotesingle s are obtained in the similar fashion except that the diagonal block entry corresponding to null vectors is replaced by $a\mathbf{I}$ where $0<a\ll 1$ is a small positive number, which indicates very low weight or confidence in null estimates (that are not shared). In essence, the central element has the full information needed to construct $\underline{\mathbf{P}}^j$\textquotesingle s and $\underline{\mathbf{R}}_{\mathbf{h}^j}$\textquotesingle s corresponding to shared estimates $\underline{\hhat}_w^j$. We illustrate how these matrices could be obtained for the example in Fig. \ref{fig:infoshare2}. Consider the central antenna $r_C{=}1$, its $|\mathcal{N}|{=}2$ direct neighbors with (shared) partial estimates given in (\ref{eq:h_Reordered}). The partial correlation and error covariance matrices associated with those estimates (shown underlined) along with that of central element are given in (\ref{eq:Rh_partial}) and (\ref{eq:P_partial}) respectively.

Based on above steps and procedures, the proposed D-LMMSE algorithm is summarized in Algorithm \ref{alg:DLMMSE1}.

\begin{align}\label{eq:Rh_partial}
\begin{aligned}
  \mathbf{R}_{\mathbf{h}^1} {=} \begin{bmatrix} \mathbf{R}_{11}&\mathbf{R}_{14}&\mathbf{R}_{12}\\
                        \mathbf{R}_{41}&\mathbf{R}_{44}&\mathbf{R}_{42}\\
                        \mathbf{R}_{21}&\mathbf{R}_{24}&\mathbf{R}_{22}\\
                   \end{bmatrix}\text{,}\:
  \underline{\mathbf{R}}_{\mathbf{h}^4} {=} \begin{bmatrix} \mathbf{R}_{44}&\mathbf{R}_{41}&\mathbf{0}\\
                        \mathbf{R}_{14}&\mathbf{R}_{11}&\mathbf{0}\\
                        \mathbf{0}&\mathbf{0}&\mathbf{I}_L\\
                   \end{bmatrix} \\
                   \text{and}\:\:
  \underline{\mathbf{R}}_{\mathbf{h}^2} = \begin{bmatrix} \mathbf{R}_{22}&\mathbf{0}&\mathbf{R}_{21}\\
                        \mathbf{0}&\mathbf{I}_L&\mathbf{0}\\
                        \mathbf{R}_{12}&\mathbf{0}&\mathbf{R}_{11}\\
                   \end{bmatrix}
\end{aligned}
\end{align}

\begin{align}\label{eq:P_partial}
\begin{aligned}
  \mathbf{P}^1 {=} \begin{bmatrix} \mathbf{P}_{11}&\mathbf{P}_{14}&\mathbf{P}_{12}\\
                        \mathbf{P}_{41}&\mathbf{P}_{44}&\mathbf{P}_{42}\\
                        \mathbf{P}_{21}&\mathbf{P}_{24}&\mathbf{P}_{22}\\
                   \end{bmatrix} \text{,}\:\:
  \underline{\mathbf{P}}^4 {=} \begin{bmatrix} \mathbf{P}_{44}&\mathbf{P}_{41}&\mathbf{0}\\
                        \mathbf{P}_{14}&\mathbf{P}_{11}&\mathbf{0}\\
                        \mathbf{0}&\mathbf{0}&a\mathbf{I}\\
                   \end{bmatrix}\\
                   \text{and}\:\:
  \underline{\mathbf{P}}^2 = \begin{bmatrix} \mathbf{P}_{22}&\mathbf{0}&\mathbf{P}_{21}\\
                        \mathbf{0}&a\mathbf{I}&\mathbf{0}\\
                        \mathbf{P}_{12}&\mathbf{0}&\mathbf{P}_{11}\\
                   \end{bmatrix}
\end{aligned}
\end{align}

\begin{algorithm}
\caption{Distributive LMMSE (D-LMMSE) algorithm}\label{alg:DLMMSE1}
 \begin{enumerate}
 \item \textbf{(Estimation)} Each antenna acting as a central element $r_C$ computes $\hhat_w^{c}$ and $\mathbf{P}^{c}$ by using $(\ref{eq:wtEstSol})$ and $(\ref{eq:estCe})$ respectively.

 \item \textbf{(Sharing)} Each antenna acting as a central element $r_C$ shares partial estimates, $\underline{\hhat}_w^c$ with its $|\mathcal{N}|$ neighbors as described in \ref{subsec:shareStep}.
 \item \textbf{(Pre-processing)} Using $\mathbf{R}_{\mathbf{h}^c}$, $\mathbf{P}^c$ from step 1 and the received (partial) information $\{\underline{\hhat}_w^j\}_{j=1}^{|\mathcal{N}|}$ in step 2, each antenna, acting as a central element $r_C$, constructs $\{\underline{\mathbf{R}}_{\mathbf{h}^j}^{-1}\}$, $\{\underline{\mathbf{P}}^j\}$, $j \in\N$.

 \item \textbf{(Update)} Each antenna acting as a central element $r_C$, updates its weighted estimate and error covariance using $(\ref{eq:estUpdate})$ and $(\ref{eq:CeUpdate})$ respectively.

 \item \textbf{(Iterate)} Repeat steps 2-4 $D$ times, where $D$ represents maximum number of iterations.

 \item \textbf{(Output)} Compute $\hhat^{c}{=}(\mathbf{P}^c)^{-1}\hhat_w^{c}$ and output the estimated CIR $\hhat_C$.
 \end{enumerate}
\end{algorithm}

\noindent\emph{Remarks:}
\begin{enumerate}
  \item The information sharing and update take place during each iteration of the algorithm such that after few iterations the information diffuses across the whole antenna array. This concept of sharing is depicted in Fig. \ref{fig:infoshare1} which shows that information diffusion process grows exponentially, resulting in fast convergence.

  \item The repetitive sharing enables each antenna in the array to utilize the observations from distant elements, thereby improving its estimate in each iteration till it converges to near optimal solution.

  \item As opposed to the central processing mechanism, the proposed sharing step is more convenient and computationally more efficient as all antennas do not communicate with each other. The collaboration takes place only among the neighboring antennas. Therefore, the complexity of proposed algorithm is significantly less than the centralized approach.

  \item Note that, the antennas share only the partial information because only selected vectors are transmitted to the neighbors which save significant amount of communication. Also, estimation step and the repetitive sharing, pre-processing and update steps require simple linear block processing and have a fixed size data structure which is well suited for real implementations.
In contrast, the memory and processing requirements for the centralized approach are even more challenging with large array dimensions.

\end{enumerate}

\subsection{Complexity Analysis}
In Table \ref{tab:comp}, we compare the  computational complexity of proposed D-LMMSE algorithm with LS, L-LMMSE and the centralized O-LMMSE algorithm in terms of multiply and add operations. The figures indicate that complexity of proposed algorithm is slightly higher (linear with BS antennas) than L-LMMSE but is significantly less than the centralized approach. For the proposed D-LMMSE algorithm, it is also worth mentioning here that, the $\mathbf{P}$ matrices  in (\ref{eq:estCe}) can be computed off-line and in parallel at all antennas as they do not depend on observations. Moreover, the computation of weighted estimates in (\ref{eq:wtEstSol}) does not involve any matrix inversion. Further, the update in (\ref{eq:estUpdate}) requires simple addition during each step of iteration, while (\ref{eq:CeUpdate}) needs one time computations of inversions $\underline{\mathbf{R}}_{\mathbf{h}^j}^{-1}$ as they do not depend on iteration index. Finally, the computation of inverse,$(\mathbf{P}^c)^{-1}$ is also required only after the convergence when each antenna outputs its final estimate.

\begin{table}[!t]
\renewcommand{\arraystretch}{1.4}
\caption{Computational Complexity}
\label{tab:comp}
\footnotesize
  \begin{center}
     \begin{tabular}{|m{1.3cm}|m{2.5cm}|m{2.1cm}|m{1.2cm}|}
     %\begin{tabular}{|c|c|c|c|}
      \hline \bfseries Algorithm & \bfseries Multiplications ($\times$) & \bfseries Additions ($+$) & \bfseries Complexity\\
      \hline LS  & $RK(L+1)$ & $R(KL-1)$ & $O(RLK)$\\
      \hline L-LMMSE & $R\big[2L^3+L^2+\newline K(L+1)\big]$ & $RL[L^{+}K{-}1]$ & $O(RL^3)$\\
      \hline {O-LMMSE} & $R\big[(L^3+1)R^2+\newline RL(L{+}K){+}K\big]{+}L^3$ & $R^2LK$ & $O(R^3L^3)$\\
      \hline D-LMMSE & $R\big[(5^3{+}1)L^3{+}2(5L)^2\newline +L(K{+}1)+5^3\big]$ & $R[D(5L)^3{+}(5L)^2\newline{+}L(K{-}1){-}D]$ & $O(R5^3L^3)$ \\
      \hline
\end{tabular}
\end{center}
\end{table}

\subsection{Choice of $D$}
\label{subsec:D}
The choice of parameter $D$ i.e., maximum number of required iterations, has a great influence on computational complexity and convergence of the proposed D-LMMSE algorithm. A trivial choice for $D$ is that it can be set to the largest dimension of the array i.e., $D{=}{\rm max}(M,G)$, which will ensure that each antenna receives information from every other antenna in the array. The aforementioned choice of $D$ would guarantee the convergence of the proposed D-LMMSE algorithm but such a high value of $D$ is very inefficient from the computational complexity point of view, particularly if the array dimensions are large. We therefore, derive a simple loose upper bound on maximum number of iterations $D$ that is much better than the trivial choice. To this end, we first note that the  total number of antennas sharing information in $D$ iterations of algorithm are $2D(D+1)+1$. Hence, in order to ensure that each antenna receives information from every other antenna in the array, we should have $2D(D+1)+1\le R$. Solving this inequality we get,
\begin{equation}\label{eq:D}
 D\le \sqrt{\frac{R}{2}-\frac{1}{4}}-\frac{1}{2}.
\end{equation}
It must be emphasised here, that the actual value of $D$ also depends on the spatial correlations among antennas. If the antennas are not very strongly correlated, then we might not gain from sharing and a small number of iterations might be sufficient. In fact, if the antennas are completely uncorrelated, then the sharing cannot improve the channel estimates as the O-LMMSE solution will converge to the L-LMMSE solution (see Section \ref{sec:LS_LMMSE_est}).
%%%%%%%%%%%%%%%%%%%%%%%%%%%%%%%%%%%%%%%%%%%%%%%
%Proof of convergence
%%%%%%%%%%%%%%%%%%%%%%%%%%%%%%%%%%%%%%%%%%%%%%%
\subsection{Convergence of D-LMMSE Algorithm}
The notion of convergence of D-LMMSE algorithm is attributed to the fact that every iteration of algorithm successively brings \textit{new information} from the neighboring tiers to the central antenna.

For example, consider the antenna array depicted in Fig. \ref{fig:infoshare1} and focus on the central antenna $r_C$.  Let $\mathbf{A}_{(i)}$ and $\mathbf{R}_{(i)}$ represent the extended data matrix and channel correlation matrix respectively, during the $i$-th iteration. Then by defining $\mathbf{I}_1\triangleq1$, we can write,
\begin{align}\label{eq:Ai_Ri}
  \mathbf{A}_{(i)} = \mathbf{I}_{i+1}\otimes\mathbf{A} \quad \text{and} \quad \mathbf{R}_{(i)} = \mathbf{R}_{array}^{(i)}\otimes\mathbf{R}_{tap}
\end{align}
where, $\mathbf{R}_{array}^{(i)}$ represents the spatial correlation matrix of the central antenna and all its neighbors up to $i$-th tier.  Further assume that, $\{\delta_l\}_{i=1}^L$ and $\{\eta_j\}_{j=1}^R$ are eigenvalues of $\mathbf{R}_{tap}$ and $\mathbf{R}_{array}$ respectively, arranged in decreasing order of magnitude. Then for $D{=}0$ (i.e., no sharing case), the resulting MSE at the central element is,
\begin{align}
  {\rm mse}_{r_C}^{(0)} &={\rm trace}\left(\mathbf{R}_{(0)}^{-1}+\mathbf{A}_{(0)}^{\rm H}\mathbf{R}_w^{-1}\mathbf{A}_{(0)}\right)^{-1}\nonumber \\
    &= \sum_{l=1}^L\left(\frac{\delta_l}{1+\rho K\delta_l}\right) \label{eq:d0}
\end{align}
which is obviously the MSE of L-LMMSE in (\ref{eq:MSESep}).
Similarly, for $D=1$ (i.e., sharing up to the first tier), $r_C$ receives information from its $|\mathcal{N}|$ neighbors (e.g., red antennas of the $1^{st}$ tier) and updates its estimate by optimal combining of neighboring estimates as described in \ref{subsec:update}. Therefore, the resulting MSE at $r_C$ can be written as,
\begin{align}
  {\rm mse}_{r_C}^1 &= \frac{1}{|\mathcal{N}^+|}\:{\rm trace}\left(\mathbf{R}_{(1)}^{-1}+\mathbf{A}_{(1)}^{\rm H}\mathbf{R}_w^{-1}\mathbf{A}_{(1)}\right)^{-1}\nonumber\:,\\
  &=\frac{1}{|\mathcal{N}^+|}\sum_{j=1}^{|\mathcal{N}^+|}\sum_{l=1}^{L}\frac{\eta_j\delta_l}{1+\rho  K\eta_j\delta_l} \:. \label{eq:d1}
\end{align}
Comparing (\ref{eq:d0}) with (\ref{eq:d1}), we note that ${\rm mse}_{r_C}^1{\le}{\rm mse}_{r_C}^0$, where the equality holds only if ${\eta_j}{=}1,\forall j$ (i.e., spatially uncorrelated channels). Proceeding similarly, it can be shown that ${\rm mse}_{r_C}^D{\le}{\rm mse}_{r_C}^{D-1}$, so that the MSE during each iteration decreases monotonically till it converges after utilizing observations from all antennas in the array.

\section{Data-aided Channel Estimation}
\label{sec:DAest}
The basic idea of data-aided channel estimation is to exploit the data sub-carriers in order to improve the initial channel estimates obtained using only the pilots. As the data aided technique does not require additional pilots, it is spectrally more efficient. Here, the pilot-based channel estimate is used for data detection, which along with the reserved pilots can significantly enhance the channel estimation. It is possible that some of the data-pilots be erroneous due to noise and channel estimation errors, while some of the other data-carriers are reliable i.e., they are likely to be decoded correctly. An important problem is how to down-select a subset of the most reliable data-carriers to be used as data-pilots.

\subsection{Reliable Carriers Selection}
Consider the received OFDM symbol at any antenna as shown in (\ref{eq:Yrt}), and let $\hhat$ and $\hat{\HH}$ be the CIR and CFR estimates obtained using pilots. Then, the tentative estimates of the data symbols are obtained by equalizing the received OFDM symbol using zero-forcing (ZF) as follows,
\begin{align}
  \hat{\XX}(k) &= \frac{\YY(k)}{\hat{\HH}(k)}, \quad\quad k\in\{1,2,\cdots,N\}\setminus\mathcal{P}\nonumber\\
  &\approx\ \XX(k) + \frac{\WW(k)}{\hat{\HH}(k)}=\XX(k) + \ZZ(k),
\end{align}
where, $\ZZ(k)$ represents the distortion on $k$-th data-carrier due to noise and channel estimation error. Given the CFR estimate, $\ZZ(k)$ can be modelled as Gaussian with zero mean and variance $\sigma_z^2{=}\hat{\HH}(k)^{-2}\sigma_w^2$. The recovery of data symbols is then performed by simple hard decisions on estimated symbols $\hat{\XX}(k)$ denoted by $\langle\hat{\XX}(k)\rangle$. Clearly, the errors in the decoding process occur due to noise as well as inaccurate channel estimates. Hence, some data-carriers would be severely effected by noise and channel perturbation errors i.e., $\mathcal{Z}(k)$ and fall outside their correct decision regions, while for some other data-carriers the  distortion is not strong enough and they are decoded correctly. All those data carriers $\hat{\XX}(k)$ which satisfy the condition $\langle\hat{\XX}(k)\rangle{=}\XX(k)$ with high probability, are termed reliable carriers.

\begin{figure}[!t]\centering
  \includegraphics[width=0.4\columnwidth]{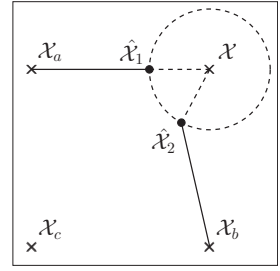}
  \caption{Concept of reliable carriers selection. Here, $\hat{\XX}(2)$ has higher probability of decoding correctly than $\hat{\XX}(1)$.}
  \label{fig:rel_concept}
\end{figure}

The proposed strategy for selecting the subset  $\mathcal{R}$ of the most reliable data-carriers, motivated by \cite{Ebrahim12}, is based on the criteria,
\begin{equation}\label{eq:Rel_exact}
  \mathbf{\mathfrak{R}}(k) {=} \frac{f_z\left(\ZZ(k){=}{\XX(k)} - \langle\hat{\XX(k)}\rangle\right)} {\sum_{m{=}1,\tiny\mathcal{A}_m\neq \langle\hat{\XX(k)}\rangle}^{M}f_z\left(\ZZ(k) {=} {\XX(k)} - \mathcal{A}_m\right)}\:,
\end{equation}
where, $f_z(.)$ is the \textit{pdf} of $\ZZ(k)$ and $\mathcal{A}_m$ represents the set of constellation alphabets. Note that the numerator in (\ref{eq:Rel_exact}) is the probability that ${\XX(k)}$ will be decoded correctly while the denominator sums the probabilities of all possible incorrect decisions due to distortion $\mathcal{Z}(k)$. The subset  $\mathcal{R}$ is formed by selecting only those data-carriers for which $\mathbf{\mathfrak{R}}(k)>1$ i.e.,
\begin{equation}\label{eq:subset_Rel}
  \mathcal{R} = \left\{k\:\:|\:\:\mathbf{\mathfrak{R}}(k)>1\right\}\:.
\end{equation}
The metric (\ref{eq:subset_Rel}) is intuitively appealing as it selects only those sub-carriers which are likely to be decoded correctly with high probability. Fig \ref{fig:rel_concept} further elaborates this idea; that even though $\hat{\XX}(1)$ and $\hat{\XX}(2)$ have the same distance from $\XX$, $\hat{\XX}(2)$ is more likely to be decoded correctly than $\hat{\XX}(1)$, as it is farther from the nearest neighbours and therefore is less likely to be decoded as any other constellation point.

\subsection{Revisiting the Estimation Step}
We now revisit the estimation step of the proposed Algorithm \ref{alg:DLMMSE1} using both the pilots and reliable carriers in order to enhance the initial estimates. Let $\mathcal{R}^r$ be the set of indices of reliable data carriers for antenna $r$, obtained in reliable carriers selection process.
Each antenna could revisit the estimation step by solving (\ref{eq:estObj}) using an extended set of indices, $\mathcal{P}\cup\mathcal{R}^r$ corresponding to pilots and reliable data carriers. To make it computationally efficient, we instead proceed by exploiting the block form of RLS to update the pilot-based estimates. Skipping the derivation, the update equations for data-aided estimation at antenna $r$ are given by,
\begin{align}
   \hhat_d^{r} &= \hhat^{r} + \mathbf{C}_e^{r}\bar{\mathbf{A}}_d^{\rm H}\mathbf{G} \left(\bar{\YY}_d-\bar{\mathbf{A}}_d\hhat^{r}\right) \label{eq:DA_hhat}\:,\\
\mathbf{C}_{ed}^{r} &= \mathbf{C}_e^{\rm r} - \mathbf{C}_e^{\rm r}\bar{\mathbf{A}}_d^{\rm H}\mathbf{G}\bar{\mathbf{A}}_d \label{eq:DA_Ce}\:,\\
\mathbf{G} &= \left(\mathbf{R}_w+\bar{\mathbf{A}}_d\mathbf{C}_e^{r}\bar{\mathbf{A}}_d^{\rm H}\right)^{-1} \label{eq:DA_Gain}\:,
\end{align}
where, $\bar{\YY}_d{=}\YY_r(\mathcal{P}\cup\mathcal{R}^r)$ is extended set of observations, $\bar{\mathbf{A}}_d{=}\begin{bmatrix}\mathbf{A}(\mathcal{P}\cup\mathcal{R}^r) & \mathbf{0}_{|\mathcal{P}\cup\mathcal{R}^r|\times |\mathcal{N}|L}\\
\end{bmatrix}$ is the extended data matrix, $\mathbf{G}$ represents the gain matrix and $\hhat^{r}$ and $\mathbf{C}_e^{r}$ are respectively the estimate and error covariance matrix obtained using only the pilots during the estimation step. The complete data-aided approach is described in Algorithm \ref{alg:DLMMSE2}.

\begin{algorithm}
\caption{Data-aided Distributive LMMSE (DAD-LMMSE) Algorithm }\label{alg:DLMMSE2}
 \begin{enumerate}
   \item Run step 1 of Algorithm \ref{alg:DLMMSE1} to get $\hhat^r$ and $\mathbf{C}_e^r$  at each antenna index $r$.
   \item Each antenna uses its CIR estimate, $\hhat^r$ to form the subset  $\mathcal{R}^r$ of the most reliable data-carriers.
   \item Update the estimates and error covariance in step (1) using (\ref{eq:DA_hhat})-(\ref{eq:DA_Ce}).
   \item Run steps (2)-(6) of Algorithm \ref{alg:DLMMSE1}, with $\mathbf{P}^r{=}(\mathbf{C}_e^r)^{-1}$ and $\hhat_w^r{=}\mathbf{P}^r\hhat^r$.
 \end{enumerate}
\end{algorithm}
% It is interesting to compare the proposed data-aided algorithm with the one given in \cite{Mudassir15}, that relies on intensive collaboration between antenna elements to enhance the data reliability. Specifically, each antenna collects reliable indices as well as equalized data from its $|\mathcal{N}|$ neighbors and then forms the set $\mathcal{R}^r$ based on the consensus strategy.
% In contrast, the proposed technique is much simple (saving computations), does not require collaboration (saving communication overhead) and yields almost identical results (see Section \ref{sec:sim}).
\section{Effect of Pilot Contamination}
\label{sec:PC}
So far, we assumed single-cell scenario where all the users have been allocated orthogonal resources for uplink channel estimation, thus the pilot observations are corrupted only by AWGN. In a multi-cell scenario, predominated by pilot contamination due to aggressive reuse of the pilots, the knowledge of the interference statistics is critical in studying the effect of pilot contamination on channel estimation techniques. Unlike the existing pilot contamination analyses, we take a stochastic geometry based approach to derive analytical expressions for interference correlation.

\subsection{Modified Network Model}
\label{subsec:NwkMod}
\begin{figure}[!t]\centering
  \includegraphics[width=0.4\columnwidth]{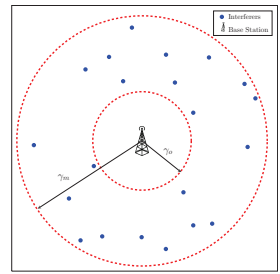}
  \caption{Realization of interferes distributed according to PPP of $\lambda{=}0.3$, $\gamma_o{=}2$m and $\gamma_m{=}5$m with BS at the origin.}
  \label{fig:int_realization}
\end{figure}

To characterise the inter-cell interference resulting from pilot contamination, we modify our previous 2-D network model of Fig. \ref{fig:cell_struct} by introducing interferes that are assumed to be distributed according to a PPP. Due to its simplicity and tractability, the PPP has been widely used in stochastic geometry for modelling of the interference in cellular networks (see \cite{Elsawy13} and references therein). Specifically, without loss of generality, we assume a single user in a reference cell of radius $\gamma_o$, communicating with the BS located at the origin $O$ in a 2-D plane. The interfering users (outside radius $\gamma_o$) are distributed over a circular region of radius $\gamma_m$  according to a homogenous PP, denoted by $\Psi$ and having intensity $\lambda$. Thus, the interfering space is an annular region with radii $\gamma_o$ and $\gamma_m$ and where the distance of $i$th interferer from BS satisfies $\gamma_o<\gamma_i<\gamma_m$. Fig. \ref{fig:int_realization} shows a realization of interferes distributed according to homogeneous PP of $\lambda{=}0.3$ with $\gamma_o{=}2$m and $\gamma_m{=}5$m.
Further, from \cite{RenzoEiD14,Haenggi09,AnumICC}, we conclude that the interference itself is not correlated across OFDM frequency tones. This makes the analysis considerably simple and tractable because each OFDM frequency tone can be treated as an independent narrow-band channel. Hence, it suffices to characterize the interference at single OFDM tone.

Consider the complex received interference at any given sub-carrier (at the BS antenna $r$) due to all interfering users, which can be represented as \cite{AnumICC},
\begin{equation}\label{eq:Iagg}
  \mathcal{I} = \sum_{i\in\Psi}\sqrt{E_x}x_ih_i
\end{equation}
where, $x_i{=}a_i{\rm exp}\{j\theta_i\}$ is the interfering symbol, $h_i{=}\gamma_i^{-\beta}\alpha_i{\rm exp}\{j\phi_i\}$ is the interfering channel, where $\beta>1$ is the pathloss exponent, $\alpha_i$ is an independent Rayleigh distributed random variable with $\Omega{=}\mathbb{E}\{\alpha_i^2\}{=}1$ and $\phi_i$ is independent random variable that is uniformly distributed over $[0,2\pi)$. The symbols $x_i$ are generated from a general bi-dimensional constellation with $M$ equiprobable symbols $\mathcal{A}_m{=}a^{(m)}{\rm exp}\{j\theta^{(m)}\}$, $m{=}1,2,\cdots,M$. We assume that all interfering users transmit with the same average energy per symbol $E_x$ and that the transmission constellation is normalized so that $\mathbb{E}\{|x_i|^2\}{=}1$. Therefore, (\ref{eq:Iagg}) can be expressed as,
\begin{equation}\label{eq:Iagg2}
  \mathcal{I} {=} \sum_{i\in\Psi\setminus O}\frac{\sqrt{E_x}a_i\alpha_i{\rm exp}\{j(\theta_i+\phi_i)\}}{\gamma_i^\beta} {=} \sum_{i\in\Psi\setminus O}\frac{\sqrt{E_x}z_i}{\gamma_i^\beta}
\end{equation}
where, $z_i=a_i\alpha_i{\rm exp}\{j(\theta_i+\phi_i)\}$.

\subsection{Interference characterization}
\label{subsec:IntCharact}
Although $\mathcal{I}$ can be completely characterized, to simplify our analysis of pilot contamination, we assume $\mathcal{I}$ to be Gaussian and thus require only the first two moments, i.e., the mean and the variance. They are given in the following lemma.

\begin{lemma}\label{lemma:int}
Using the network model of \ref{subsec:NwkMod}, the mean and variance of interference $\mathcal{I}$ is,
\begin{equation}\label{eq:muI}
  \mu_{\tiny{\mathcal{I}}} = \mathbb{E}\{\mathcal{I}\}=0
\end{equation}
and
\begin{align}\label{eq:varI}
  \sigma_{\tiny{\mathcal{I}}}^2 &= \mathbb{E}\{|\mathcal{I}|^2\} \nonumber\\
  &=\pi\lambda (\beta-1)^{-1}\mathbb{E}\{|x|^2\}E_x\Omega \left(\frac{1}{\gamma_o^{2\beta-2}}-\frac{1}{\gamma_m^{2\beta-2}}\right)
\end{align}
respectively.
\end{lemma}

\begin{IEEEproof}
  See Appendix \ref{append:mu_varI}.
\end{IEEEproof}

Although (\ref{eq:varI}) is derived by considering that the interference space is annular, it can be extended for an infinite interference space with a protection region of $\gamma_o$ by taking the limit as $\gamma_m\rightarrow\infty$ yielding,
\begin{equation}\label{eq:varI2}
  \sigma_{\tiny{\mathcal{I}}}^2 =\pi\lambda \gamma_o^2(\beta-1)^{-1}\mathbb{E}\{|x|^2\} \left(\frac{E_x\Omega}{\gamma_o^{\beta-1}}\right)\:\:.
\end{equation}

\subsection{Effect of PC on MSE Performance}\label{subsec:PC_MSE}
The knowledge of interference statistics at single OFDM frequency tone, obtained through lemma \ref{lemma:int}, allows us to evaluate the aggregate interference correlation over all OFDM tones and/or across the whole BS antenna array using known channel statistics. Consider the received OFDM symbol at $r$th BS antenna, after omitting the index $\mathcal{P}$,
\begin{align}\label{eq:YrPC}
  \YY_{r} &= \mathbf{A}\mathbf{h}_{r} + \II_r + \WW_{r} \nonumber \\
  &= \mathbf{A}\mathbf{h}_{r} + \EE_{r}
\end{align}
where, $\II_r$ is the interference at antenna $r$ of BS due to pilot contamination and $\EE_r$ is the interference term which captures the effect of both pilot contamination and the noise. Due to independence of noise and pilot contamination terms, each having a zero mean, the correlation matrix of $\EE_r$ is  $\mathbf{R}_{\tiny\EE_r}{=} \mathbf{R}_{\tiny\II_r}{+}\mathbf{R}_{w}$. Now, using the interference power (or variance) at each OFDM sub-carrier from lemma \ref{lemma:int}, the interference correlation matrix $\mathbf{R}_{\tiny\II_r}$ across OFDM tones can be easily obtained as $\mathbf{R}_{\tiny\II_r}{=}\sigma_{\tiny{\mathcal{I}}}^2\mathbf{A} \mathbf{R}_{tap}\mathbf{A}^{\rm H}$, where we assumed that all user channels (both desired and interfering) have identical correlations (as in section \ref{sec:sysMod}) and use the same pilots, which is the worst case scenario from pilot contamination perspective. Similarly, in the multi-antenna case, based on system model of (\ref{eq:Y_pilots}), the interference correlation matrix for the whole BS array can be obtained as $\mathbf{R}_{\tiny\EE} {=} \mathbf{R}_{\tiny\II}{+}\mathbf{R}_w$, with $\mathbf{R}_{\tiny\II}{=}\sigma_{\tiny{\mathcal{I}}}^2\acute{\mathbf{A}} \mathbf{R}_{\mathbf{h}}\acute{\mathbf{A}}^{\rm H}$. Using these interference correlations, we can derive the MSE expressions for LS, L-LMMSE and O-LMMSE algorithms in the presence of noise and pilot contamination by replacing the noise covariance matrix $\mathbf{R}_w{=}\sigma^2\mathbf{I}$ with matrix $\mathbf{R}_{\tiny\EE_r}$ or $\mathbf{R}_{\tiny\EE}$ in the MSE expressions already obtained in section \ref{sec:LS_LMMSE_est}. The results are presented in following theorems.

\begin{theorem}\label{MSE:LS_PC}
For the system model described in section \ref{sec:sysMod} and pilot contamination as characterised in section \ref{sec:PC}, the MSE expression for LS estimation algorithm of section \ref{subsec:LS_est} under both AWGN and pilot contamination is given by,
\begin{equation}\label{eq:MSE_LSPC}
  {\rm MSE}^{\rm(LS)}=\frac{RL}{\rho K}+ R\sigma_{\tiny{\mathcal{I}}}^2\:{\rm trace}(\Lambda)\:,
\end{equation}
where, $\sigma_{\tiny{\mathcal{I}}}^2$ is given in (\ref{eq:varI}) and  $\Lambda$ is a diagonal matrix with eigenvalues of $\mathbf{R}_{tap}$ spread along the diagonal and all users are assumed to have similar channel characteristics.
\end{theorem}
\begin{IEEEproof}
See Appendix \ref{append:LS_PC}.
\end{IEEEproof}
Theorem \ref{MSE:LS_PC}, shows that MSE is composed of two terms. The first term due to AWGN can be suppressed by increasing the number of pilot tones but the second term due to pilot contamination cannot be reduced by adding more pilots and even persists at high SNR (i.e., $\rho\rightarrow\infty$).

\begin{theorem}\label{MSE:LLMMSE_PC}
For the system model described in section \ref{sec:sysMod} and pilot contamination as characterised in section \ref{sec:PC}, the MSE expression for L-LMMSE estimation algorithm presented in section \ref{subsec:LLMMSE_est} under both AWGN and pilot contamination is given by,
\begin{equation}\label{eq:MSE_LLMMSEPC}
  {\rm MSE}^{\rm(L)}=R\sum_{i=1}^L\frac{\delta_i\left(1 + \rho K\delta_i\sigma_{\tiny{\mathcal{I}}}^2\right)}{1+\rho K\delta_i+\rho K\delta_i\sigma_{\tiny{\mathcal{I}}}^2}\:,
\end{equation}
where, $\sigma_{\tiny{\mathcal{I}}}^2$ is given in (\ref{eq:varI}), $\delta_i$ are the eigenvalues of  $\mathbf{R}_{tap}$ and all users are assumed to have similar channel characteristics.
\end{theorem}
\begin{IEEEproof}
Replace $\mathbf{R}_w$ with $\mathbf{R}_w+\mathbf{R}_{\tiny \mathcal{I}_r}$ in MSE expression (\ref{eq:MSESep}), then invoking the eigenvalue decomposition (EVD) of $\mathbf{R}_{tap}$, follow the steps of Theorem \ref{MSE:LS_PC} given in Appendix \ref{append:LS_PC}. We skip the detailed proof due to its similarity to Theorem \ref{MSE:LS_PC}.
\end{IEEEproof}
Note that (\ref{eq:MSE_LLMMSEPC}) reduces to MSE expression for AWGN (given in (\ref{eq:MSESep})) had there been no pilot contamination. At high SNR (i.e. $\rho\gg1$), when there essentially remains only the effect of pilot contamination, the MSE expression (\ref{eq:MSE_LLMMSEPC}) reduces to,
\begin{equation}\label{eq:MSE_LLMMSEPC2}
{\rm MSE}^{\rm(L)}\:\:\overset{\tiny{high\:SNR}}\longrightarrow \:\:R\left(\frac{\sigma_{\tiny{\mathcal{I}}}^2} {1+\sigma_{\tiny{\mathcal{I}}}^2}\right){\rm trace}(\mathbf{\Lambda})\:,
\end{equation}
which shows that MSE is independent of number of pilots and that  LMMSE is more robust to pilot contamination compared to LS.
\begin{theorem}\label{MSE:OLMMSE_PC}
For the system model described in section \ref{sec:sysMod} and pilot contamination as characterised in section \ref{sec:PC}, the MSE expression for O-LMMSE estimation algorithm presented in section \ref{subsec:OLMMSE_est} under both AWGN and pilot contamination is given by,
\begin{equation}\label{eq:MSE_OLMMSEPC}
  {\rm MSE}^{\rm(O)}=\sum_{j=1}^{R}\sum_{i=1}^L\frac{\mu_j\delta_i \left(1 + \rho K\mu_j\delta_i\sigma_{\tiny{\mathcal{I}}}^2\right)}{1+\rho K\mu_j\delta_i+\rho K\mu_j\delta_i\sigma_{\tiny{\mathcal{I}}}^2}\:\:,
\end{equation}
where, $\sigma_{\tiny{\mathcal{I}}}^2$ is given in (\ref{eq:varI}), $\mu_j$ and $\delta_i$ are the eigenvalues of $\mathbf{R}_{array}$ and $\mathbf{R}_{tap}$ respectively, and all users are assumed to have similar channel characteristics.
\end{theorem}
\begin{IEEEproof}
See Appendix \ref{append:OLMMSE_PC}.
\end{IEEEproof}
Note that (\ref{eq:MSE_OLMMSEPC}) reduces to the MSE expression for AWGN given in (\ref{eq:MSEOpt}) in absence of pilot contamination. Again observe that, under the assumption of high SNR, when the effect of pilot contamination predominates AWGN, the MSE expression in (\ref{eq:MSE_OLMMSEPC}) simplifies to,
\begin{equation}\label{eq:MSE_OLMMSEPC2}
  {\rm MSE}^{\rm(O)}\overset{\tiny{high\:SNR}}\longrightarrow \left(\frac{\sigma_{\tiny{\mathcal{I}}}^2} {1+\sigma_{\tiny{\mathcal{I}}}^2}\right){\rm trace}(\mathbf{R}_{array}){\rm trace}(\mathbf{\Lambda}).
\end{equation}
This indicates that MSE depends strongly on interference power and is independent of number of pilots $K$. Since ${\rm trace}(\mathbf{R}_{array})\le R$, the O-LMMSE seems to be more robust to pilot contamination compared to both LS and L-LMMSE. The MSE expression also gives us clue that effect of pilot contamination can be minimized by exploiting the spatial correlations and by optimizing the BS antenna array design.

Above theorems quantify the effect of pilot contamination on MSE performance of channel estimation in terms of interference power (or variance) which in turn depends on different parameters  described in lemma \ref{lemma:int}. The MSE performance against various parameters will be numerically analysed through simulations.

\section{Simulation Results}
\label{sec:sim}
We adopt the channel model in (\ref{eq:Rht}) with spatial correlation matrix given in (\ref{eq:Rs_kron}) whose parameters are: $\phi{=}\pi/3$ (mean horizontal AoD in radians), $\theta{=}3\pi/8$ (mean vertical AoD in radians), $\sigma{=}\pi/12$ (standard deviation of horizontal AoD) and $\xi{=}\pi/36$ (standard deviation of vertical AoD). The channel tap correlation matrix follows an exponentially decaying PDP, $\mathbb{E}\{|h_r{(\tau)}|^2\}{=}e^{-\tau}$, while rest of the parameters are given in the Table \ref{tab:sim_par}, where $\nu$ represents the carrier frequency wavelength in meters. It is also assumed that receiver has the knowledge of channel correlations.
\begin{table}[!t]
\renewcommand{\arraystretch}{1.4}
\caption{Parameters for simulation}\label{tab:sim_par}
\footnotesize
  \begin{center}
     \begin{tabular}{|l|c|}
      \hline \textbf{Parameter} & \textbf{Value} \\
      \hline  Array Size $(M\times G)$ & 10 $\times$10\\
      \hline  Array element spacing $d_x, d_y$ & 0.3$\nu$, 0.5$\nu$\\
      \hline  Number of OFDM sub-carriers $(N)$ & 256\\
      \hline  Number of pilots $(K)$ & 32\\
      \hline  Signal constellation modulation & 4/16/64 -- QAM\\
      \hline  Channel length $(L)$ & 8\\
      \hline
  \end{tabular}
 \end{center}
\end{table}

To assess the performance of different algorithms we use the following MSE performance criterion:
\begin{equation}\label{MSE_def}
 MSE = \frac{1}{\Theta}\sum_{i=1}^{\Theta}\|\mathbf{h}^i-\hat{\mathbf{h}}^i\|^2
\end{equation}
where, $\mathbf{h}^i$ and $\hat{\mathbf{h}}^i$ are true and estimated CIR vectors (at the $i$th trial) respectively, each of size $RL\times 1$ and $\Theta$ represents the total number of trials. We used $\Theta{=}100$ in our simulations.

We conduct five different experiments to study the performance of our proposed approach and compare it with the three methods i.e., LS, L-LMMSE and O-LMMSE described earlier in Section \ref{sec:LS_LMMSE_est}. We also perform  experiments to validate our analysis and study the impact of pilot contamination on all these methods.
\subsection{Experiment 1: How many iterations ($D$)?}
\label{NPC_results}
\begin{figure}[!t]\centering
 \subfigure[]{
  \includegraphics[width=0.465\linewidth]{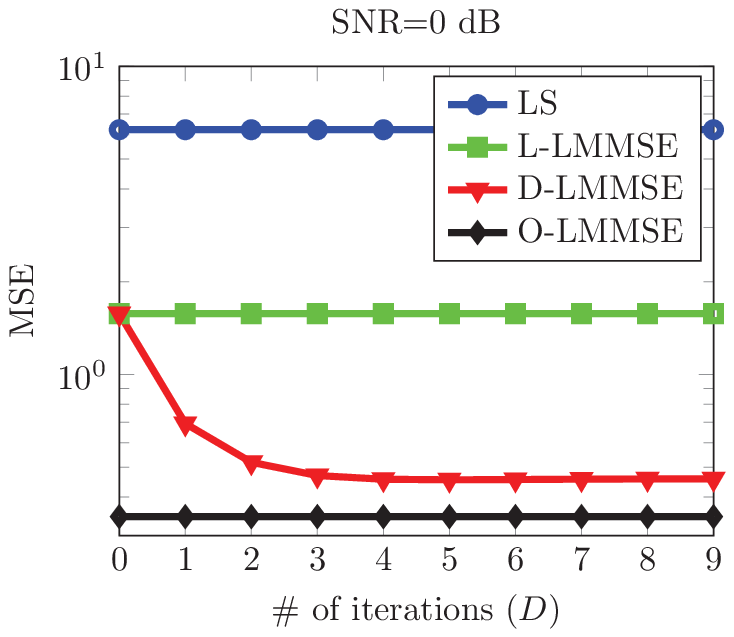}
  \label{fig:mse_iter1}
 }
\subfigure[]{
  \includegraphics[width=0.465\linewidth]{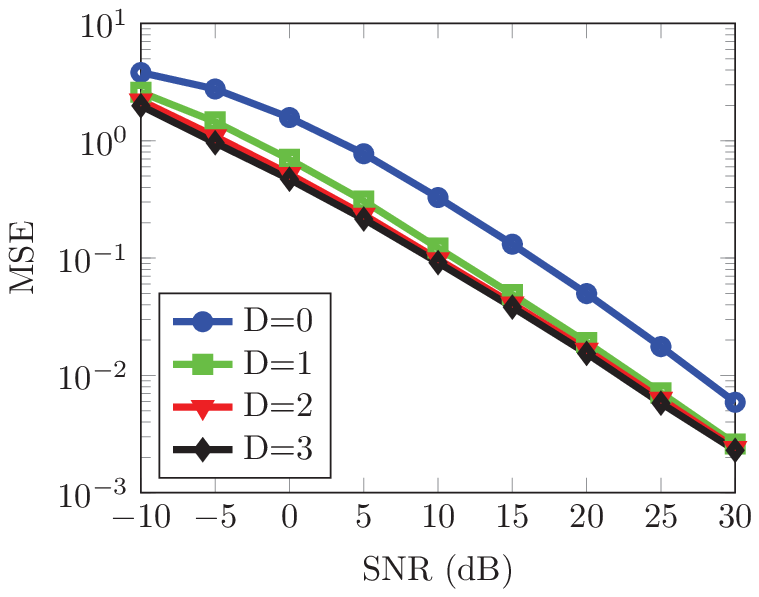}
  \label{fig:mse_iter2}
  }
  \caption{Number of iterations ($D$) required to achieve the convergence of distributed algorithm.}
  \label{fig:howManyiters}
\end{figure}

In this experiment we are interested in finding the number of iterations, required for convergence of the proposed distributed LMMSE algorithm. We plot the MSE of proposed D-LMMSE algorithm (red curve) against the parameter $D$ (i.e., number of iterations) in Fig. \ref{fig:mse_iter1}. The SNR was fixed at $0$ dB. The MSE values of other algorithms, which do not depend on parameter $D$, are also shown. It can be seen that the proposed algorithm converges very closely to the optimal in 3 iterations. Note that, when the antennas do not collaborate (i.e., $D{=}0$), the MSE of distributed algorithm coincides with that of L-LMMSE because no information sharing takes place. As the information from neighbors comes in during the next few iterations, the MSE decays exponentially until it converges to near optimal solution. Fig. \ref{fig:mse_iter2} also suggests that there would be hardly any improvement in MSE for $D>3$.

\subsection{Experiment 2: MSE Performance in AWGN}
\begin{figure}[!t]\centering
  \includegraphics[width=0.65\columnwidth]{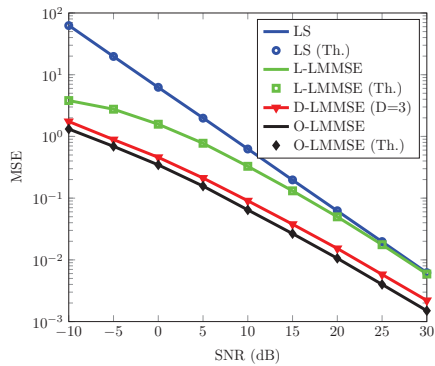}
  \caption{MSE performance of different algorithms in white Gaussian noise.}
  \label{fig:mse_snr1}
\end{figure}

In this experiment, we compare the MSE performance of different algorithms in the presence of AWGN using the parameters in Table \ref{tab:sim_par}. The results given in Fig. \ref{fig:mse_snr1}, show that O-LMMSE performs better than both LS and L-LMMSE in terms of MSE as it is able to utilize the antenna spatial correlations. As shown, the proposed D-LMMSE algorithm\ (Algorithm \ref{alg:DLMMSE1})  achieves near optimal results in just 3 iterations. The analytical MSE expressions given in Section \ref{sec:LS_LMMSE_est}, for LS, L-LMMSE and O-LMMSE under AWGN are also plotted with legends (Th.), which agree with simulation results.

\begin{figure}[!t]\centering
 \subfigure[]{
  \includegraphics[width=0.46\columnwidth]{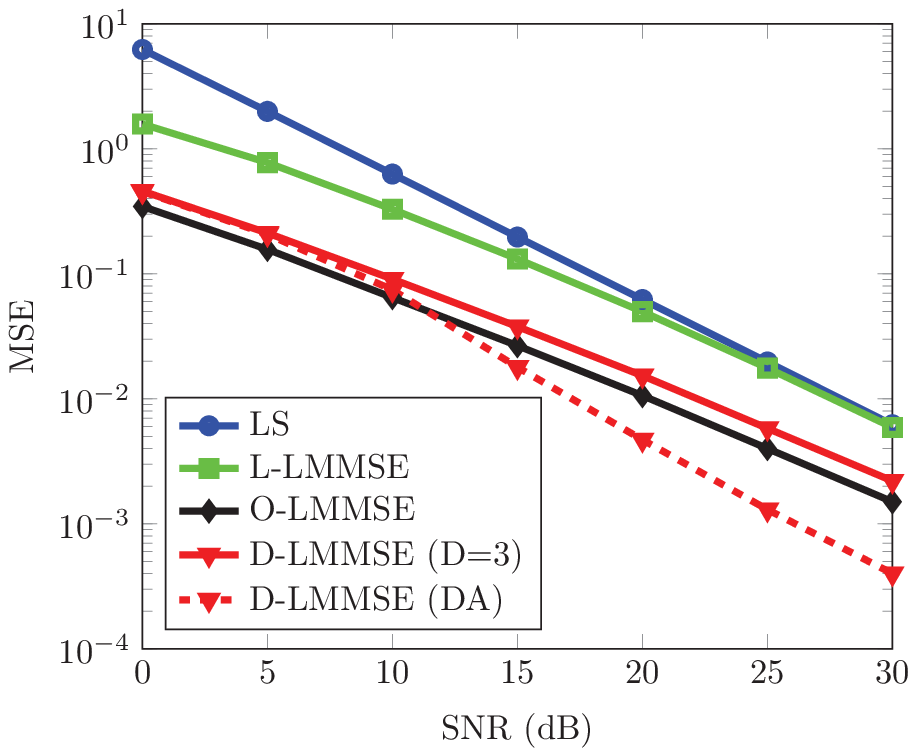}
  \label{fig:mse_snrRel}
 }
\subfigure[]{
  \includegraphics[width=0.46\columnwidth]{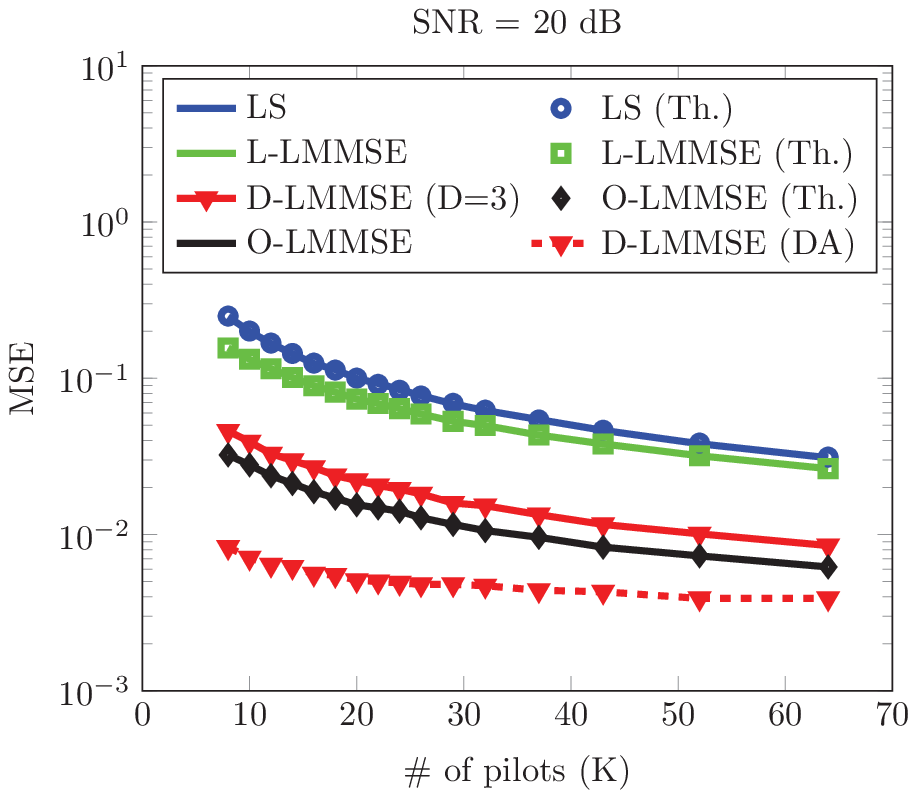}
  \label{fig:mse_pilots}
  }
  \caption{MSE performance comparison of data-aided D-LMMSE algorithm with pilot-based techniques in white Gaussian noise.}
  \label{fig:data_aided}
\end{figure}

Fig. \ref{fig:mse_snrRel} shows the MSE performance of proposed data-aided algorithm (DAD-LMMSE in Algorithm \ref{alg:DLMMSE2}) against other pilot-based algorithms. It is obvious that data-aided approach has the best performance compared to all others and that the effect of using reliable carriers is more pronounced at higher  SNR. Fig. \ref{fig:mse_pilots} demonstrates the MSE behaviour of different algorithms with varying number of pilots $K$ with SNR fixed at 20 dB. As is shown, increasing the pilot tones yields better estimation performance but this comes at the cost of lower spectral efficiency. The data-aided algorithm however, is able to achieve the best performance even for a small number of pilot tones.

\subsection{Experiment 3: Mean and variance of interference}
\label{subsec:muVar_lamda}
\begin{figure}[!t]\centering
  \includegraphics[width=0.78\linewidth]{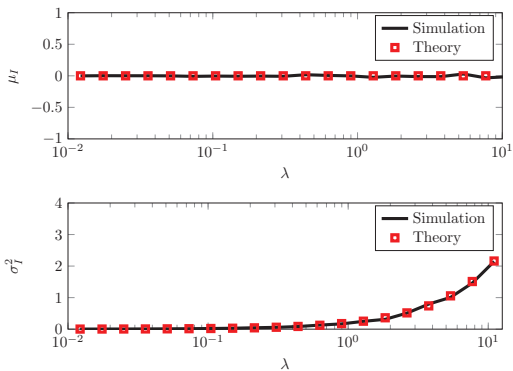}
  \caption{Mean and variance of interference at single OFDM sub-carrier as a function of $\lambda$.}
  \label{fig:muVar_lamda}
\end{figure}

This experiment aims to validate the mean and variance of the interference given in Lemma \ref{lemma:int}. In order to mimic the setup described in Section \ref{subsec:NwkMod}, we use single antenna BS and assume that CIRs from each user to the BS has a uniform PDP. Further, we assume that BS is located at the origin, the desired user at a distance of 1m from BS while interfering users are distributed in a region of radius 5m and with a protection region of $\gamma_o{=}2$m according to a PPP with density $\lambda$ and pathloss exponent $\beta{=}2$. All users communicate with BS using OFDM with $N{=}256$, $L{=}8$ and $K{=}32$ identical pilot symbols drawn from a $4$-QAM constellation. Fig. \ref{fig:muVar_lamda} compares the mean and variance of interference observed on single OFDM carrier (randomly picked) due to simulated sources with expressions given in Lemma \ref{lemma:int}, as a function of $\lambda$. The results indicate a close match between simulation and theory.

\subsection{Experiment 4: MSE Performance under AWGN and Pilot Contamination}
\label{subsec:effect_of_PC}
\begin{figure}[!t]\centering
 \subfigure[]{
  \includegraphics[width=0.45\linewidth]{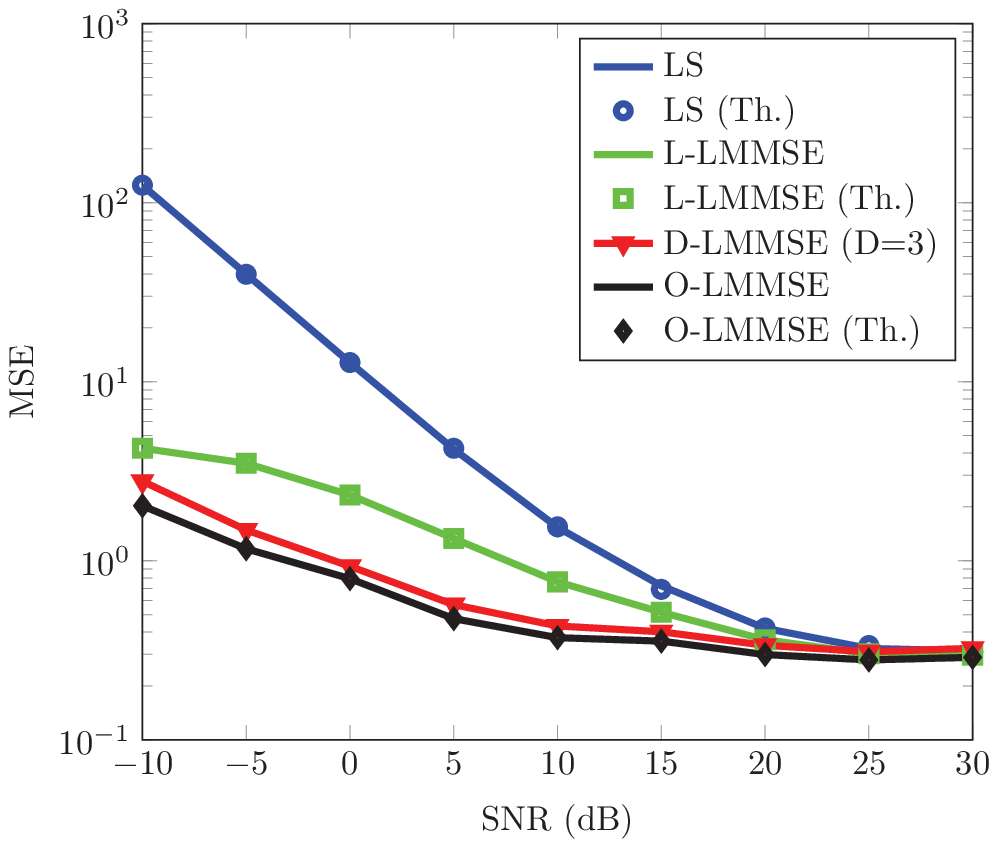}
  \label{fig:mse_snrPC}
 }
\subfigure[]{
  \includegraphics[width=0.45\linewidth]{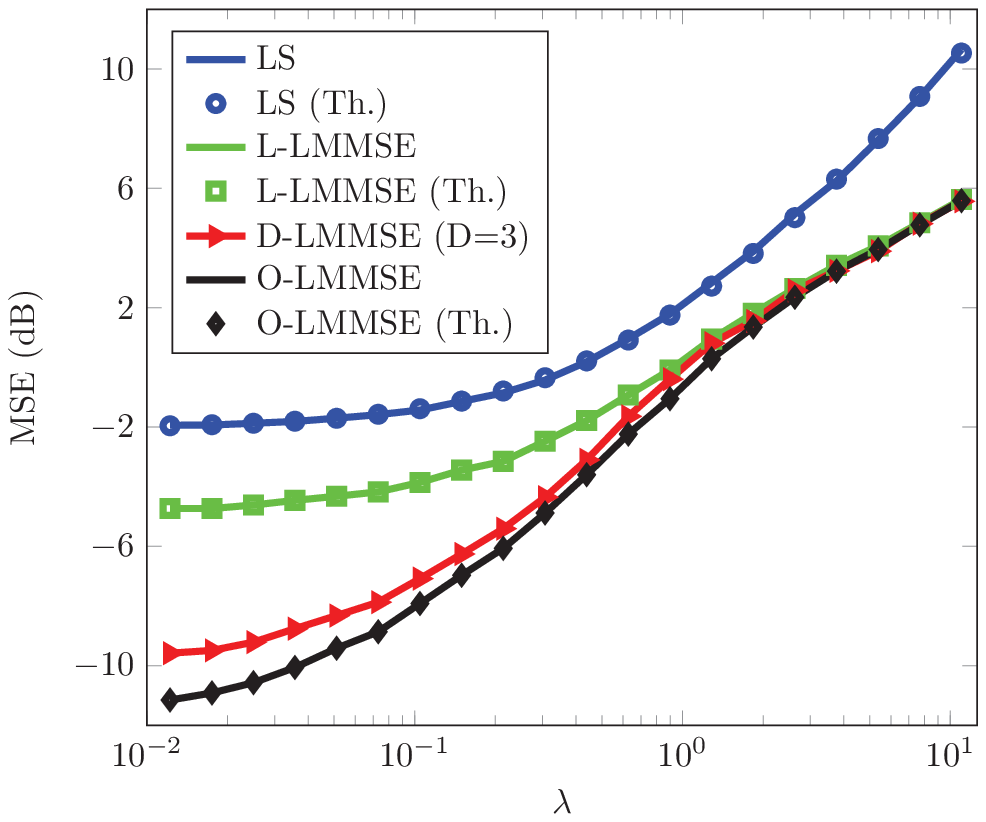}
  \label{fig:mse_lamda}
  }
  \caption{Effect of pilot contamination on MSE performance \subref{fig:mse_snrPC} MSE as a function of SNR for $\lambda{=}0.1$ \subref{fig:mse_lamda} MSE as a function of $\lambda$ for SNR fixed at $10$ dB.}
  \label{fig:PC_results}
\end{figure}

In this experiment we study the MSE performance of different algorithms in presence of both AWGN and pilot contamination. For simulations, we use the parameters given in Table \ref{tab:sim_par} with the interfering users distributed according to a PPP of $\lambda{=}0.1$ and pathloss $\beta{=}2$. The desired user is assumed 1m away from BS located at origin while the interfering users are distributed in circular region of radius 5m with protection region of $\gamma_o{=}2$ m. In Fig. \ref{fig:mse_snrPC}, the simulated MSE performance of different algorithms is compared over a wide range of SNR with the analytical expressions given in Theorems \ref{MSE:LS_PC}, \ref{MSE:LLMMSE_PC} and \ref{MSE:OLMMSE_PC} (see Section \ref{subsec:PC_MSE}). From Fig. \ref{fig:mse_snrPC}, note that all MSE curves decrease with increasing SNR in lower range but reach an error floor at higher SNR. This is in stark contrast to AWGN case (see Fig. \ref{fig:mse_snr1}), where the MSE always decreases with increasing SNR. This shows that pilot contamination persists even at higher SNR and its effect on MSE is more severe than AWGN.

We present similar analysis in Fig. \ref{fig:mse_lamda}, where the MSE is plotted as a function of $\lambda$ with SNR fixed at $10$ dB. It is obvious that all algorithms perform well for small values of $\lambda$. However when $\lambda$ increases, the interference due to pilot contamination dominates AWGN, thus severely degrading the performance as indicated by a sharp increase in MSE curves. Note that LMMSE channel estimation is more robust to pilot contamination than simple LS based channel estimation.
Also observe a close match between simulation and theoretical analysis, shown in Fig. \ref{fig:PC_results}, over a wide range of $\lambda$.
\subsection{Experiment 5: Computational Complexity}
\begin{figure}[!t]\centering
  \includegraphics[width=0.78\linewidth]{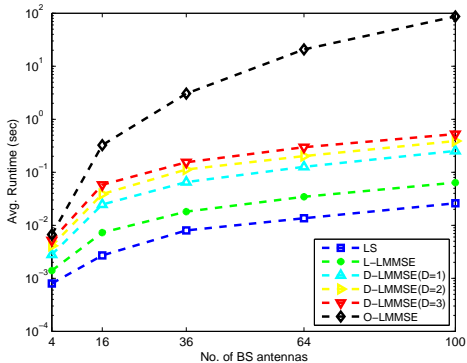}
  \caption{Average runtime of various algorithms.}
  \label{fig:sim_time}
\end{figure}

In this experiment we compare the average runtime of various algorithms that can be regarded as a measure of computational complexity. Fig. \ref{fig:sim_time} shows the average runtime with increasing number of BS antennas under the default simulation parameters of Table \ref{tab:sim_par}. It is clear that computational requirements for proposed D-LMMSE algorithm, with different values of parameter $D$,  grow at much slower pace than that of the O-LMMSE algorithm as the number of BS antenna increases. Further, in terms of memory requirements and communication overhead (not shown here), the advantages of D-LMMSE are even more tangible.

\section{Conclusion}
\label{sec:concl}
Channel estimation is a challenging problem in massive MIMO systems as the conventional techniques applicable to MIMO systems cannot be employed owing to an exceptionally large number of unknown channel coefficients. We proposed a distributed algorithm that attains near optimal solution at a significantly reduced complexity by relying on coordination among antennas. To reduce the pilots overhead, the distributed LMMSE algorithm is extended using data-aided estimation based on reliable carriers. To gain insight into the effect of pilot contamination on channel estimation performance, we used the stochastic geometry to obtain the aggregated interference power and then based on this, we derived MSE expressions for different algorithms under AWGN and pilot contaminated scenarios. The derived expressions were verified using simulation results. Extending the obtained results to analyzing the system throughput under pilot contamination remains open for future work.
\appendices
\section{Mean and variance of interference}
\label{append:mu_varI}

The mean of $\mathcal{I}$ can be determined as follows,
\begin{align*}\label{eq:App_muI}
  \mu_{\tiny{\mathcal{I}}} &= \mathbb{E}\{\mathcal{I}\}=\mathbb{E}\left\{\sum_{i\in\Psi }\frac{\sqrt{E_x}z_i}{\gamma_i^\beta}\right\}\nonumber \\
  &= \mathbb{E}_{\Psi}\left\{\sum_{i\in\Psi}\frac{\sqrt{E_x}\:\mathbb{E}_z\{z_i\}} {\gamma_i^\beta}\right\}\nonumber \\
&\overset{(a)}{=} \sqrt{E_x}\mathbb{E}\{z_i\}\int_{\mathbb{R}^2}\frac{1} {r^\beta}r dr d\theta=0
\end{align*}
where, $\overset{(a)}{=}$ results from Campbell's theorem \cite{chiu2013} and then the fact, $\mathbb{E}\{z_i\}=0$ yields the zero mean. Similarly, the variance of interference can be computed as follows,
\begin{align*}%\label{eq:App_varI}
  \sigma^2_{\tiny{\mathcal{I}}} &= \mathbb{E}\{|\mathcal{I}|^2\}\\
  &= \mathbb{E}_{\Psi}\left\{\mathbb{E}_{z}\sum_{i\in\Psi}\frac{\sqrt{E_x}z_i} {\gamma_i^\beta} \sum_{j\in\Psi}\frac{\sqrt{E_x}z_j^*}{\gamma_j^\beta}\right\}\\
&\overset{(a)}{=} \mathbb{E}_{\Psi}\left\{\sum_{i\in\Psi}\frac{E_x\mathbb{E}_{z}\{|z_i|^2\}} {\gamma_i^{2\beta}} \right\}\\
&\overset{(b)}{=} \lambda E_x\mathbb{E}\{|z_i|^2\}\int_0^{2\pi}\int_{\gamma_o}^{\gamma_m} \frac{1}{r^{2\beta}}r dr d\theta \\
&\overset{(c)}{=} \pi\lambda (\beta-1)^{-1}E_x\Omega\mathbb{E}\{|x|^2\}\left( \frac{1}{\gamma_o^{2\beta-2}}-\frac{1}{\gamma_m^{2\beta-2}}\right)
\end{align*}
where, $\overset{(a)}{=}$ is due to the fact that $z_i$ are independent SS random variables, in $\overset{(b)}{=}$ we employed Campbell's theorem and in $\overset{(c)}{=}$ we used the result $\mathbb{E}\{|z_i|^2\}=\mathbb{E}\{a_i^2\alpha_i^2\}=\Omega\mathbb{E}\{|x|^2\}$, where we note that $a_i$ and $\alpha_i$ are independent random variables, which completes the proof.

\section{Proof of Theorem \ref{MSE:LS_PC}}
\label{append:LS_PC}

By replacing $\mathbf{R}_w$ with $\mathbf{R}_w+\mathbf{R}_{\tiny \mathcal{I}_r}$ in MSE expression of (\ref{eq:mseSepLS}), we obtain
\begin{align*}
  {\rm mse}_r^{\rm ls} &= {\rm trace}\left(\mathbf{A}^{\rm H}\left(\mathbf{R}_w+\mathbf{R}_{\tiny \mathcal{I}_r}\right)^{-1}\mathbf{A}\right)^{-1} \\
&= {\rm trace}\left(\mathbf{A}^{\rm H}\left(\mathbf{R}_w+\sigma_{\tiny \mathcal{I}}^2\mathbf{A} \mathbf{R}_{tap}\mathbf{A}^{\rm H}\right)^{-1}\mathbf{A}\right)^{-1}\\
&\overset{(a)}= {\rm trace}\Bigl(\mathbf{A}^{\rm H}\mathbf{R}_w^{-1}\mathbf{A}-\sigma_{\tiny \mathcal{I}}^2\mathbf{A}^{\rm H}\mathbf{R}_w^{-1}\mathbf{A}\bigl(\mathbf{R}_{tap}^{-1}\\ &\phantom{={\rm trace}\Bigl(} %make space for cont. line
+\sigma_{\tiny \mathcal{I}}^2\mathbf{A}^{\rm H}\mathbf{R}_w^{-1}\mathbf{A}\bigr)^{-1}\mathbf{A}^{\rm H}\mathbf{R}_w^{-1}\mathbf{A}\Bigr)^{-1}
\end{align*}
where, $\overset{(a)}=$ follows from matrix inversion lemma. Now, using the EVD of the channel correlation matrix $\mathbf{R}_{tap}=\mathbf{Q}\mathbf{\Lambda}\mathbf{Q}^{\rm H}$ and the fact that $\mathbf{A}^{\rm H}\mathbf{R}_w^{-1}\mathbf{A}=\frac{KE_x}{\sigma_w^2}\mathbf{I}_L$ we obtain,
\begin{align}
  {\rm mse}_r^{\rm ls} &\overset{(b)}= {\rm trace}\Biggl(\frac{KE_x}{\sigma_w^2}\mathbf{I}_L-\sigma_{\tiny \mathcal{I}}^2\left(\frac{KE_x}{\sigma_w^2}\right)^2\Bigl(\mathbf{\Lambda}^{-1}\nonumber\\
&\phantom{= {\rm trace}\Biggl(}+\frac{\sigma_{\tiny \mathcal{I}}^2KE_x}{\sigma_w^2}\mathbf{I}_L\Bigr)^{-1}\Biggr)^{-1}\nonumber\\
&{=} \sum_{i=1}^L\left(\frac{KE_x}{\sigma_w^2}{-}\sigma_{\tiny \mathcal{I}}^2\left(\frac{KE_x}{\sigma_w^2}\right)^2 \left(\delta_i^{-1}{+}\frac{\sigma_{\tiny \mathcal{I}}^2KE_x}{\sigma_w^2}\right)^{-1}\right)^{-1}
\end{align}
where, $\overset{(b)}=$ follows from the property that ${\rm trace}\left(\mathbf{Q}\mathbf{R}\mathbf{Q}^{\rm H}\right)={\rm trace}(\mathbf{R})$ if $\mathbf{Q}$ is unitary. After simple algebraic manipulations, the term inside the summation simplifies to $\frac{\sigma_w^2L}{KE_x}+\sigma_{\tiny \mathcal{I}}^2\sum_{i=1}^L\delta_i$, which completes the proof.

\section{Proof of Theorem \ref{MSE:OLMMSE_PC}}
\label{append:OLMMSE_PC}
Under both AWGN and pilot contamination, we replace $\mathbf{R}_w$ with $\mathbf{R}_{\tiny \EE}=\mathbf{R}_w+\sigma_{\tiny{\mathcal{I}}}^2\acute{\mathbf{A}} \mathbf{R}_{\mathbf{h}}\acute{\mathbf{A}}^{\rm H}$ to get,
\begin{align*}
  {\rm MSE}^{\rm (O)} &= {\rm trace}\left(\mathbf{R}_{\mathbf{h}}^{-1}{+} \acute{\mathbf{A}}^{\rm H}\left(\mathbf{R}_w{+}\sigma_{\tiny{\mathcal{I}}}^2\acute{\mathbf{A}} \mathbf{R}_{\mathbf{h}}\acute{\mathbf{A}}^{\rm H}\right)^{-1}\acute{\mathbf{A}}\right)^{-1}\\
&\overset{(a)}= {\rm trace}\Biggl(\mathbf{R}_{\mathbf{h}}^{-1}+ \acute{\mathbf{A}} \mathbf{R}_{\mathbf{h}}\acute{\mathbf{A}}^{\rm H}-\sigma_{\tiny{\mathcal{I}}}^2\acute{\mathbf{A}} \mathbf{R}_{\mathbf{h}}\acute{\mathbf{A}}^{\rm H}\Bigl(\mathbf{R}_{\mathbf{h}}^{-1}\\
&\phantom{= {\rm trace}\Biggl(}+\sigma_{\tiny{\mathcal{I}}}^2\acute{\mathbf{A}} \mathbf{R}_{\mathbf{h}}\acute{\mathbf{A}}^{\rm H}\Bigr)^{-1}\acute{\mathbf{A}} \mathbf{R}_{\mathbf{h}}\acute{\mathbf{A}}^{\rm H}\Biggr)^{-1}
\end{align*}
 where $\overset{(a)}=$ follows from matrix inversion lemma. Using the properties of kronecker product, it can be shown that $\acute{\mathbf{A}} \mathbf{R}_{\mathbf{h}}\acute{\mathbf{A}}^{\rm H}=\frac{KE_x}{\sigma_w^2}(\mathbf{I}_{R}\otimes\mathbf{I}_L)$. Further, the channel correlation matrix $\mathbf{R}_{\mathbf{h}}=\mathbf{R}_{array}\otimes\mathbf{R}_{tap}$ can be decomposed as $\mathbf{R}_{\mathbf{h}}=(\mathbf{V}\otimes\mathbf{Q})(\mathbf{S}\otimes\mathbf{\Lambda}) (\mathbf{V}\otimes\mathbf{Q})^{\rm H}$, where we introduced the EVDs, $\mathbf{R}_{array}=\mathbf{V}\mathbf{S}\mathbf{V}^{\rm H}$ and $\mathbf{R}_{tap}=\mathbf{Q}\mathbf{\Lambda}\mathbf{Q}^{\rm H}$. Incorporating these results in $\overset{(a)}=$ yields,

\begin{align*}
  {\rm MSE}^{\rm (O)} &\overset{(b)}{=} {\rm trace}\Biggl(\mathbf{S}^{-1}{\otimes}\mathbf{\Lambda}^{-1}{+}\frac{KE_x}{\sigma_w^2}(\mathbf{I}_{R}\otimes\mathbf{I}_L){-} \sigma_{\tiny{\mathcal{I}}}^2\left(\frac{KE_x}{\sigma_w^2}\right)^2\\
&\phantom{= {\rm trace}\Biggl(}\cdot\left(\mathbf{S}^{-1}\otimes\mathbf{\Lambda}^{-1}{+}\frac{\sigma_{\tiny{\mathcal{I}}}^2KE_x}{\sigma_w^2} (\mathbf{I}_{R}\otimes\mathbf{I}_L)\right)^{-1}\Biggr)^{-1}\\
&\overset{(c)}= \sum_{j=1}^{R}\sum_{i=1}^L\Biggl(\frac{1}{\mu_j\delta_i}+\frac{KE_x}{\sigma_w^2}- \sigma_{\tiny{\mathcal{I}}}^2\left(\frac{KE_x}{\sigma_w^2}\right)^2 \\
&\phantom{= \sum_{j=1}^{R}\sum_{i=1}^L\Biggl(}\cdot\left(\frac{1}{\mu_j\delta_i}+ \frac{\sigma_{\tiny{\mathcal{I}}}^2KE_x}{\sigma_w^2}\right)^{-1}\Biggr)^{-1}
\end{align*}
where, $\overset{(b)}=$ follows from property,  ${\rm trace}\left(\mathbf{Q}\mathbf{R}\mathbf{Q}^{\rm H}\right){=}{\rm trace}(\mathbf{R})$ when $\mathbf{Q}$ is unitary and $\overset{(c)}=$ is due to the diagonal nature of the matrix inside the ${\rm trace}$ operator, where $\mu_j$ and $\delta_i$ represent the eigenvalues of matrices $\mathbf{R}_{array}$ and $\mathbf{R}_{tap}$ respectively. After some algebraic manipulations, $\overset{(c)}=$ simplifies to the result given in Theorem \ref{MSE:OLMMSE_PC}.

\bibliographystyle{IEEEtran}
\bibliography{IEEEabrv,Biblio_main}
%\end{footnotesize}

\end{document}